\documentclass[aps,prl,twocolumn,letterpaper,superscriptaddress,amsmath,amssymb,floatfix,nofootinbib]{revtex4-2}
\usepackage{graphicx,bm,tikz}
\usepackage[colorlinks=true,linkcolor=blue!50!black,citecolor=blue!50!black,urlcolor=blue!50!black]{hyperref}
\usetikzlibrary{arrows.meta,decorations.pathmorphing,calc}
\newcommand{\MapRfour}{1.14}
\newcommand{\MapRsix}{0.82}
\newcommand{\MapReight}{0.58}
\newcommand{\MapTfour}{0.56}
\newcommand{\MapTsix}{0.44}
\newcommand{\MapTeight}{0.35}

\newcommand{\low}{^{\mathrm{low}}}
\newcommand{\Tr}{\mathrm{Tr}}
\begin{document}
\makeatletter\let\qfi@addcontentsline\addcontentsline\def\addcontentsline#1#2#3{}\makeatother   %

\title{Certifying Nonlocal Spin Quantum Entanglement from Local Tunneling Spectra}

\author{Payman Kazemikhah}
\email{payman@princeton.edu}
\affiliation{Laboratory for Topological Quantum Matter and Advanced Spectroscopy, Department of Physics, Princeton University, Princeton, NJ, USA.}
\affiliation{Department of Chemistry, Princeton University, Princeton, NJ, USA.}
\author{Ruolun Zhang}
\affiliation{Laboratory for Topological Quantum Matter and Advanced Spectroscopy, Department of Physics, Princeton University, Princeton, NJ, USA.}
\author{Xiangyu Luo}
\affiliation{Laboratory for Topological Quantum Matter and Advanced Spectroscopy, Department of Physics, Princeton University, Princeton, NJ, USA.}
\author{Byunghoon Kim}
\affiliation{Laboratory for Topological Quantum Matter and Advanced Spectroscopy, Department of Physics, Princeton University, Princeton, NJ, USA.}
\affiliation{Department of Chemistry, Princeton University, Princeton, NJ, USA.}
\author{Yu-Peng Zhu}
\affiliation{Laboratory for Topological Quantum Matter and Advanced Spectroscopy, Department of Physics, Princeton University, Princeton, NJ, USA.}
\author{Md. Shafayat Hossain}
\affiliation{Quantum Materials Intelligence \& Nano Devices (QMIND) Lab, Department of Materials Science and Engineering, University of California Los Angeles (UCLA), CA, USA.}
\author{M. Zahid Hasan}
\email{mzhasan@princeton.edu}
\affiliation{Laboratory for Topological Quantum Matter and Advanced Spectroscopy, Department of Physics, Princeton University, Princeton, NJ, USA.}

\begin{abstract}
Quantum entanglement is intrinsically nonlocal, whereas tunneling spectroscopy measures one site at a time. We show that this apparent mismatch can be overcome: local tunneling spectra can certify spin entanglement without coherent control or absolute intensity calibration. An exact spin sum rule turns the high-bias conductance into an internal normalization, allowing the inelastic spin-flip weight to place a lower bound on the local quantum Fisher information. Total-spin conservation then fixes the otherwise inaccessible intersite fluctuation, converting two local spectra into a bound on a collective entanglement witness. We incorporate coupling to the substrate through a spectroscopic bound on spin leakage, while a hierarchy of thresholds distinguishes entanglement between the spins from entanglement with their environment. Including instrumental broadening and noise, the resulting certificate holds at a stated confidence level. These results establish a route for extracting nonlocal quantum information from equilibrium local spectroscopy.
\end{abstract}
\maketitle

Entanglement is both a resource for quantum technologies and a defining feature of many-body quantum states~\cite{guhne2009,pezze2018}. In a material, however, it is difficult to establish directly. The degrees of freedom of interest are embedded in a larger system, and strong correlations alone do not prove that two chosen spins are entangled. A useful experiment should therefore do more than reproduce a microscopic model's predictions. It should provide a quantitative certificate of entanglement and identify which local degrees of freedom share it.

Quantum Fisher information (QFI) provides a direct connection between measurable fluctuations and entanglement. For thermal states, QFI can be obtained from the dynamical susceptibility~\cite{braunstein1994,pezze2009,hyllus2012,toth2012,hauke2016,shitara2016}, and neutron scattering has used this relation to witness entanglement in quantum magnets~\cite{mathew2020,scheie2021,laurell2021}. These measurements probe collective fluctuations averaged over many sites. Scanning tunneling microscopy (STM) reaches the opposite limit~\cite{yin2021}. Inelastic electron tunneling spectroscopy (IETS) resolves spin excitations of individual atoms, molecules, and engineered spin clusters~\cite{heinrich2004,hirjibehedin2006,fernandezrossier2009,lorente2009,ternes2015,ternes2017,choi2019,mishra2021}. This raises a natural question. Can quantum entanglement between two spins be certified from spectra measured locally, one site at a time?

The difficulty is that the information needed for such a certificate is not obviously present in a local spectrum. Entanglement witnesses depend on correlations between different spins, while IETS measures fluctuations of a single spin~\cite{guhne2009,krammer2009,wiesniak2005,malla2024}. The spectral intensity also contains unknown tunneling amplitudes, so it has no immediate absolute normalization. In addition, the spins exchange angular momentum with substrate electrons and can become entangled with that environment. A large local spin-flip signal can therefore arise even when the two observed spins are not entangled. Existing proposals for atomic-scale entanglement detection avoid some of these difficulties by using coherent control with electron-spin-resonance STM~\cite{baumann2015,yang2019,wang2023,delcastillo2023,broekhoven2024}. Whether ordinary equilibrium tunneling spectra contain enough information has remained unclear.

Here we show that they do because the spectra obey strong physical constraints. A spin of fixed length $S$ has a fixed total fluctuation weight, $S(S+1)$. Once all spin-flip channels are open, the high-bias conductance provides an internal reference for this weight. The inelastic step can then be converted into a calibration-free lower bound on the local QFI [Eq.~\eqref{eq:prop1}]. The missing intersite information follows from total-spin conservation. For a dimer, conservation fixes the covariance between the two spins and converts two local QFI bounds into a lower bound on the collective QFI [Eq.~\eqref{eq:closure}].

Coupling to the substrate weakens this conservation law, but in a measurable way. We describe the loss of spin conservation by a leakage parameter $\epsilon$ and bound it from the excitation linewidth. This allows the collective QFI to remain a rigorous lower bound even for an open system. We then derive thresholds that distinguish ordinary product states, correlations mediated through the environment, and entanglement of the two spins themselves [Eqs.~\eqref{eq:tiers} and \eqref{eq:t0}]. We include instrumental broadening and measurement noise explicitly, so the final result is an entanglement certificate with a stated confidence level.

The broader point is that local measurements can contain enough information to establish a nonlocal quantum property when exact sum rules fix the available fluctuation weight and conservation laws constrain how that weight is shared. In this setting, equilibrium spin-excitation spectroscopy serves as an atomically resolved entanglement witness without coherent driving or absolute intensity calibration.

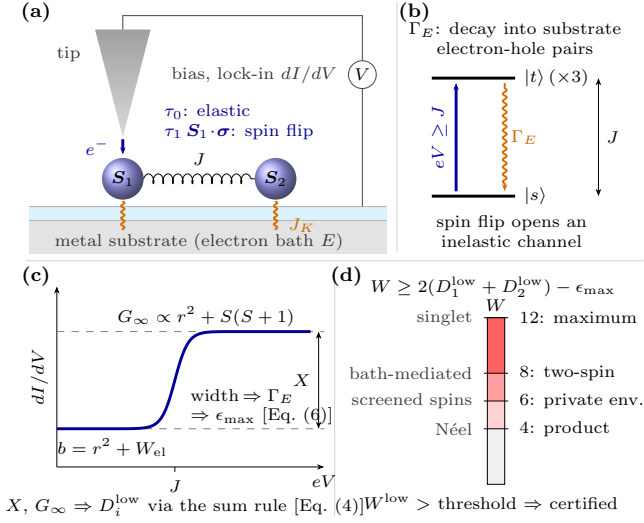
\begin{figure}[t]
\centering
\resizebox{\columnwidth}{!}{%
\begin{tikzpicture}[x=1cm,y=1cm,every node/.style={font=\scriptsize},>={Stealth[length=3pt]}]
\draw[gray!35] (5.45,3.55)--(5.45,7.15); \draw[gray!35] (0.1,3.42)--(9.0,3.42); \draw[gray!35] (4.55,-0.25)--(4.55,3.32);
\node[font=\footnotesize\bfseries] at (0.25,7.0) {(a)};
\fill[gray!22] (0.15,3.52) rectangle (5.3,4.0);
\fill[cyan!12] (0.15,4.0) rectangle (5.3,4.2);
\draw[gray!60] (0.15,4.0)--(5.3,4.0) (0.15,4.2)--(5.3,4.2);
\node[gray!45!black] at (2.6,3.7) {metal substrate (electron bath $E$)};
\foreach \x/\l in {1.5/1,3.7/2}{
  \draw[decorate,decoration={snake,amplitude=0.7pt,segment length=2.5pt},orange!85!black,thick] (\x,4.28)--(\x,3.86);
  \shade[ball color=blue!35] (\x,4.6) circle (0.3);
  \node at (\x,4.6) {$\bm S_{\l}$};}
\node[orange!80!black] at (4.08,3.93) {$J_K$};
\draw[decorate,decoration={coil,aspect=0.5,amplitude=2pt,segment length=4pt}] (1.8,4.6)--(3.4,4.6);
\node at (2.6,4.88) {$J$};
\shade[top color=gray!75,bottom color=gray!30] (1.1,6.75)--(1.9,6.75)--(1.5,5.2)--cycle;
\node[gray!40!black] at (0.72,6.4) {tip};
\draw[->,blue!65!black,very thick] (1.5,5.17)--(1.5,4.95);
\node[blue!65!black,anchor=east] at (1.42,5.06) {$e^-$};
\node[anchor=west,align=left,blue!55!black] at (1.98,5.42) {$\tau_0$: elastic\\ $\tau_1\,\bm S_1\!\cdot\!\bm\sigma$: spin flip};
\draw[gray!55!black] (1.5,6.75)--(1.5,6.95)--(4.95,6.95)--(4.95,4.2);
\draw[gray!55!black,fill=white] (4.95,6.1) circle (0.2); \node at (4.95,6.1) {$V$};
\node[anchor=east,gray!45!black] at (4.7,6.1) {bias, lock-in $dI/dV$};
\node[font=\footnotesize\bfseries] at (5.7,7.0) {(b)};
\node[align=center] at (7.15,6.6) {$\Gamma_E$: decay into substrate\\electron-hole pairs};
\draw[very thick] (5.95,4.35)--(7.15,4.35) node[right] {$|s\rangle$};
\draw[very thick] (5.95,6.05)--(7.15,6.05) node[right] {$|t\rangle\,(\times3)$};
\draw[<->] (8.35,4.35)--(8.35,6.05) node[midway,right] {$J$};
\draw[->,blue!65!black,very thick] (6.3,4.42)--(6.3,5.98) node[midway,sloped,above] {$eV\ge J$};
\draw[->,orange!85!black,thick,decorate,decoration={snake,amplitude=0.8pt,segment length=3pt,pre length=2pt,post length=3pt}] (7.0,5.98)--(7.0,4.42);
\node[orange!80!black,anchor=west] at (6.98,5.2) {$\Gamma_E$};
\node[align=center] at (7.1,3.85) {spin flip opens an\\inelastic channel};
\node[font=\footnotesize\bfseries] at (0.25,3.2) {(c)};
\draw[->] (0.55,0.45)--(4.4,0.45) node[below] {$eV$};
\draw[->] (0.55,0.45)--(0.55,3.05);
\node[rotate=90] at (0.3,1.75) {$dI/dV$};
\draw[dashed,gray] (0.55,1.0)--(4.4,1.0) (0.55,2.4)--(4.4,2.4);
\draw[very thick,blue!60!black,samples=120,domain=0:2.15,variable=\u] plot ({0.55+1.7*\u},{1.0+0.7*(1+tanh(max(-4,min(4,9*(\u-1)))))});
\draw (2.25,0.45)--(2.25,0.38) node[below] {$J$};
\node at (1.35,0.73) {$b=r^2+W_{\rm el}$};
\node at (2.7,2.62) {$G_\infty\propto r^2+S(S+1)$};
\draw[<->] (4.32,1.0)--(4.32,2.4) node[midway,left] {$X$};
\node[anchor=west,align=left] at (2.36,1.32) {width$\,\Rightarrow\Gamma_E$\\$\Rightarrow\epsilon_{\max}$ [Eq.~(6)]};
\node at (2.4,-0.1) {$X$, $G_\infty\Rightarrow D_i^{\rm low}$ via the sum rule [Eq.~(4)]};
\node[font=\footnotesize\bfseries] at (4.72,3.2) {(d)};
\node[anchor=west] at (4.95,3.05) {$W\ge2(D_1^{\rm low}+D_2^{\rm low})-\epsilon_{\max}$};
\fill[gray!12] (6.75,0.2) rectangle (7.0,1.0);
\fill[red!18] (6.75,1.0) rectangle (7.0,1.4);
\fill[red!38] (6.75,1.4) rectangle (7.0,1.8);
\fill[red!62] (6.75,1.8) rectangle (7.0,2.6);
\draw (6.75,0.2) rectangle (7.0,2.6);
\node at (6.875,2.78) {$W$};
\foreach \w/\t/\ex in {4/product/N\'eel, 6/private env./screened spins, 8/two-spin/bath-mediated, 12/maximum/singlet}{
  \draw (6.7,0.2+0.2*\w)--(7.05,0.2+0.2*\w);
  \node[anchor=west] at (7.1,0.2+0.2*\w) {\w: \t};
  \node[anchor=east,gray!50!black] at (6.65,0.2+0.2*\w) {\ex};}
\node at (6.8,-0.05) {$W^{\rm low}>$ threshold $\Rightarrow$ certified};
\end{tikzpicture}}
\caption{\textbf{From local tunneling spectra to an entanglement certificate.}
(a)~An STM tip probes one site of an exchange-coupled spin dimer with interaction $J$. Each spin couples to the metallic substrate via Kondo exchange $J_K$. Tunneling contains a spin-conserving channel $\tau_0$ and a spin-exchange channel $\tau_1\bm S_i\!\cdot\!\bm\sigma$.
(b)~For $eV\ge J$, spin-flip tunneling excites the singlet to the triplet. The excitation decays into substrate electron-hole pairs at rate $\Gamma_E$.
(c)~The excitation produces a step in $dI/dV$. The fixed total spin weight $S(S+1)$ makes the high-bias plateau $G_\infty$ an internal normalization for the inelastic weight $X$. Their ratio gives the calibration-free lower bound $D_i^{\rm low}$ on the local QFI [Eq.~\eqref{eq:prop1}]. The step width determines $\Gamma_E$ and bounds the leakage $\epsilon_{\max}$ [Eq.~\eqref{eq:kondo}].
(d)~Total-spin conservation constrains how the two local fluctuations are correlated. Equation~\eqref{eq:closure} therefore converts the two local bounds into a lower bound on the collective QFI $W$, including the correction for spin leakage. The thresholds $W=4$, $6$, and $8$ distinguish product states, spins correlated only through their environments, and entanglement of the two spins themselves, respectively.}
\label{fig:concept}
\end{figure}

Consider $N$ spins $\bm S_i$ of spin $S$ coupled to each other and to an environment $E$ in a joint state $\rho$. The environment includes substrate electrons, nuclear spins and unresolved neighbors. The collective QFI summed over the three spin axes is $W(\bm w)=\sum_{a}F_Q[\rho,\sum_iw_iS_i^a]=\bm w^{T}K\bm w$. It is determined by the QFI matrix $K=\sum_aK^a$ of the operators $S_i^a$~\cite{braunstein1994}. The diagonal elements $D_i=K_{ii}$ are the local QFIs. The QFI of the total cluster spin, $\epsilon=\bm 1^TK\bm 1=\sum_aF_Q[\rho,S^a_{\rm tot}]$, vanishes when $\rho$ commutes with $\bm S_{\rm tot}$, for example for an isotropic cluster without environment. Otherwise $\epsilon$ measures how strongly the total spin leaks into $E$. For a dimer we write $W\equiv W(1,-1)$.

To second order in tunneling [Fig.~\ref{fig:concept}(a)], with energy-independent amplitudes and unpolarized electrodes, the conductance over site $i$ is~\cite{appelbaum1967,fernandezrossier2009,ternes2015}
\begin{equation}
G(V)=\kappa\Big[\tau_0^2+\tau_1^2\!\int\! dE\,S_i(E)\,\theta_T(eV,E)\Big].
\label{eq:forward}
\end{equation}
Here $S_i(E)=\sum_a\sum_{mn}p_n|\langle m|S_i^a|n\rangle|^2\delta(E-E_m+E_n)$ is the local spin structure factor, and $\theta_T(\omega,E)=\Phi_T'(\omega-E)+\Phi_T'(-\omega-E)$ is the thermal step, with $\Phi_T(x)=x/(1-e^{-x/T})$ at electron temperature $T$. The prefactors $\kappa$, $\tau_0$ and $\tau_1$ are unknown. For a thermal state at spin temperature $T_s=1/\beta$, $S_i$ obeys detailed balance, $S_i(-E)=e^{-\beta E}S_i(E)$, and the sum rule $\int S_i=S(S+1)$. The local QFI then follows from the spectrum~\cite{hauke2016},
\begin{equation}
D_i=4\int_0^\infty\! dE\,w_\beta(E)\,S_i(E),\quad w_\beta(E)=\tanh\tfrac{\beta E}{2}\big(1-e^{-\beta E}\big).
\label{eq:thermal}
\end{equation}
Equation~\eqref{eq:forward} agrees with explicit spin-resolved rates (Supplemental Material (SM)~\cite{SM}, Sec.~S3). It holds exactly for a polarized tip on an unpolarized substrate if $\rho$ is invariant under spin rotations. Polarization of both electrodes only renormalizes $\tau_0^2\to\tau_0^2(1+\Pi)$ and $\tau_1^2\to\tau_1^2(1-\Pi/3)$, with $\Pi=P_tP_s\,\hat{\bm n}\cdot\hat{\bm m}$.

\begin{figure}[t]
\includegraphics[width=\columnwidth]{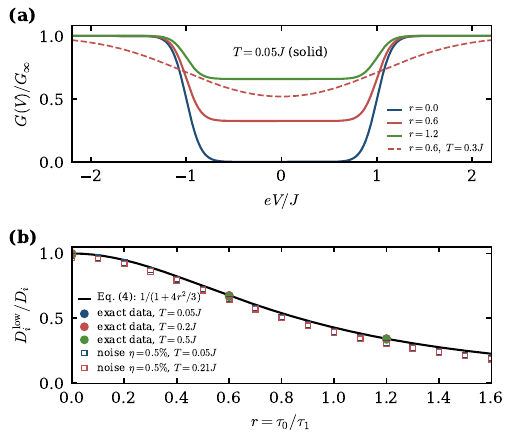}
\caption{\textbf{Calibration-free bound on the local QFI.}
(a) Normalized conductance of the isolated dimer at $T=0.05J$ for $r=\tau_0/\tau_1=0$, 0.6 and 1.2 (solid) and at $T=0.3J$ with $r=0.6$ (dashed), for a lock-in amplitude of $0.02J$. (b) $D_i\low/D_i$ versus $r$. The line shows Eq.~\eqref{eq:prop1}. Filled symbols are optimized bounds for exact spectra ($h=T/32$ at $0.05J$ and $T/64$ otherwise). Open symbols are bounds for noise $\eta=0.5\%$, 181 bias points and a joint confidence level of 95\%.}
\label{fig:theory}
\end{figure}

The sum rule has a simple physical content. The weight $S(S+1)$ is shared between elastic spin scattering and the spin-flip excitations. Above the highest excitation all of it contributes, so the plateau conductance is $\kappa[\tau_0^2+S(S+1)\tau_1^2]$ and serves as an internal ruler for the unknown tunneling strength. In units of $\kappa\tau_1^2$ the conductance reads $g(V)=b+\int\Theta(eV;E)\,dX(E)$. Here $X$ is the spectral weight at $E>0$, and $\Theta(\omega;E)=\theta_T(\omega,E)+e^{-\beta E}\theta_T(\omega,-E)$ includes its detailed-balance partner. The baseline $b=r^2+W_{\rm el}$ combines the spin-conserving channel, $r=\tau_0/\tau_1$, with the elastic spin weight. The ratio
\begin{equation}
R[b,X]=\frac{4S(S+1)\int w_\beta\,dX}{b+\int(1+e^{-\beta E})\,dX}
\label{eq:R}
\end{equation}
is independent of $\kappa$ and $\tau_1$. By Eq.~\eqref{eq:thermal} its numerator is $S(S+1)D_i$, and by the sum rule its denominator is $r^2+S(S+1)$. Hence
\begin{equation}
R=\frac{D_i}{1+r^2/S(S+1)}\le D_i .
\label{eq:prop1}
\end{equation}
Any lower bound on $R$ over the spectra compatible with the data is therefore a calibration-free lower bound on $D_i$. The only unknown left is the spin-conserving channel $\tau_0$, which adds to the plateau without adding spin fluctuations. It can only lower $R$, and the factor $1+r^2/S(S+1)$ is the price of not knowing it [Figs.~\ref{fig:concept}(c) and \ref{fig:theory}(b)]. Spectral weight above the measured window acts as a constant background, which the ratio compensates. It costs at most $8e^{-\beta E_{\rm cut}}$ times the missing weight, where $E_{\rm cut}$ is the upper end of the window (SM Sec.~S4).

The matrix $K$ is positive semidefinite, with diagonal $D_i$ and total sum $\epsilon$. For a dimer its off-diagonal element, the correlation between the fluctuations of the two spins, is therefore fixed at $K_{12}=(\epsilon-D_1-D_2)/2$. If the total spin is conserved ($\epsilon=0$), the two spins fluctuate exactly in opposite directions. The collective QFI $W=D_1+D_2-2K_{12}$ then obeys the parallelogram relation
\begin{equation}
W=2(D_1+D_2)-\epsilon\ \ge\ 2(D_1\low+D_2\low)-\epsilon_{\max}.
\label{eq:closure}
\end{equation}
Two local spectra and an upper bound on the leakage thus bound the collective QFI. For a dimer the optimal bound on $\lambda_{\max}(K)$ follows in closed form (SM Sec.~S5). For $N$ sites $\lambda_{\max}(K)\ge(\sum_iD_i-\epsilon/N)/(N-1)$. A single measured site gives $W\ge(2\sqrt{D_1}-\sqrt{\epsilon})^2$ for $D_1\ge\epsilon$. For Heisenberg chains of up to 12 sites, however, local bounds certify only the product threshold at low temperature (SM Sec.~S8). Beyond dimers, intersite information is needed.

For any Hamiltonian, $F_Q[\rho_\beta,G]\le\beta\langle[G,[H,G]]\rangle_\beta$ (SM Sec.~S6). Anisotropic couplings therefore produce a leakage bounded by their strength. For Kondo exchange $J_K$ with substrate electrons at $T\ll J$, second-order perturbation theory relates $\epsilon$ to the linewidth $\Gamma_E$ of the singlet-triplet excitation~\cite{loth2010},
\begin{equation}
\epsilon_0=\frac{12}{\pi}\frac{\Gamma_E}{J}\,\Lambda\!\left(\frac{W_b}{J}\right),\qquad \Lambda(x)=\ln(1+x)-\frac{x}{1+x},
\label{eq:kondo}
\end{equation}
where $W_b$ is the bandwidth ($\Lambda\approx4.53$ at $W_b/J=250$). Exact diagonalization with narrow discretized bands confirms Eq.~\eqref{eq:kondo} to 6\% for $J_K\le0.1J$ (SM Sec.~S6). For the wide bands of metals ($W_b/J=100$--1000), full-density-matrix numerical renormalization-group (NRG) calculations~\cite{wilson1975,bulla2008,weichselbaum2007} give a somewhat larger leakage, $\epsilon=\xi\epsilon_0$. Here $\xi=1.2$--1.6 for $\nu J_K\le0.05$ at $T\le0.1J$, and $\xi$ reaches 2.0 at $T=0.5J$, with $\nu$ the density of states per spin (SM Sec.~S8). These values hold for spins coupled to independent bands. When both spins couple to one band, with overlap $S$ between their local orbitals, $\xi$ grows with $S$ already at $T\ll J$. It reaches 1.7 at $S=0.6$ and 2.2 at $S=0.9$ for $\nu J_K=0.05$, and 3.2 at $S=0.9$ and $T=0.3J$ (SM Sec.~S8). The bound $\epsilon_{\max}$ must then use the largest $\xi$ among the plausible overlaps. With $\Gamma_E$ the measured step width, $\epsilon_{\max}=\xi\epsilon_0$ is thus obtained spectroscopically. For strongly overlapping orbitals, the substrate also induces a ferromagnetic coupling of order $J_K^2\nu$ between the spins. At full overlap NRG finds a singlet for $2(W_b/J)(\nu J_K)^2\le0.5$. For values of 2 and larger the ground state is an underscreened triplet with $W=4$, and no threshold is passed (SM Sec.~S8).

Because $D_i$ and $\epsilon$ refer to the full state, a large $W$ need not imply that the two spins are entangled with each other. Exchange with environment modes can mimic intersite correlations. We therefore distinguish three situations (SM Sec.~S2). If spins and environment are in a separable state, each spin fluctuates like a spin coherent state, and $W\le4$. Each spin may instead be entangled with its own part of the environment, as for Kondo screening by independent baths. If the two composite subsystems are then unentangled, $W\le6$. Finally, for every state the fluctuations of $\bm S_1-\bm S_2$ are limited by the singlet weight $a=\langle\psi^-|\rho_{12}|\psi^-\rangle$ of the reduced two-spin state,
\begin{equation}
W\le4\langle(\bm S_1-\bm S_2)^2\rangle=4(1+2a).
\label{eq:tiers}
\end{equation}
Hence $W>8$ implies $a>\frac12$ and thus entanglement of the spins themselves~\cite{horodecki1999}, with concurrence $C(\rho_{12})\ge(W-8)/4$~\cite{wootters1998,verstraete2002}. For $S\ge1$ the first two bounds coincide at $4S(S+1)|\bm w|^2$, and the last becomes $W>8S(2S+1)$. All thresholds are reached. A N\'eel product state gives $W=4$. Two spins, each in a singlet with a private partner, give $W=6$ with a maximally mixed $\rho_{12}$. The state $(|{\uparrow\downarrow}\rangle|e_0\rangle+|{\downarrow\uparrow}\rangle|e_1\rangle)/\sqrt2$ gives $W=8$ with a separable $\rho_{12}$. The criterion $W>6$ makes no assumption about how the substrate is shared. The criterion $W>8$ certifies entanglement of the spins themselves, which is the resource relevant for spin qubits~\cite{wang2023}.

\begin{figure}[t]
\includegraphics[width=\columnwidth]{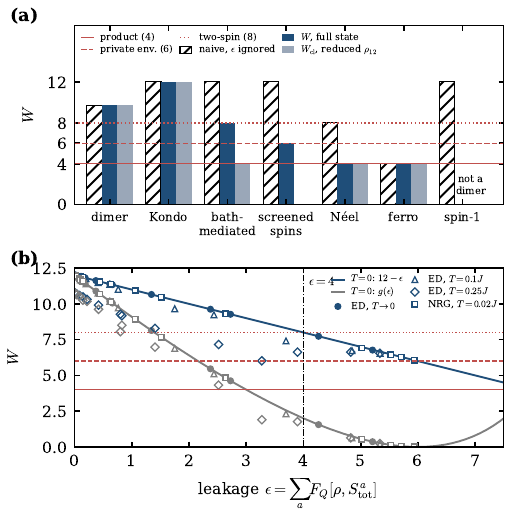}
\caption{\textbf{States that mimic entanglement and Kondo screening.}
(a) Naive estimate $2(D_1+D_2)$ (hatched), full-state $W$ (blue) and reduced $W_{\rm cl}$ (gray) for the isolated dimer ($T=0.3J$), the Kondo-coupled dimer ($J_K=0.1J$, $T\to0$), the bath-mediated state, privately screened spins, a N\'eel product state, a ferromagnetic dimer and a misidentified spin-1 ion. Lines mark the thresholds. (b) $W$ (blue) and $W_{\rm cl}$ (gray) of the Kondo-coupled dimer versus $\epsilon$. Circles are from exact diagonalization with discretized baths ($J_K/J=0.05$--3.5), filled for $T\to0$ and open for $T=0.1J$ and $0.25J$. Squares are from NRG with continuous bands at $T=0.02J$. Lines show Eq.~\eqref{eq:t0}.}
\label{fig:controls}
\end{figure}

At low temperature the criteria become sharp. Suppose the ground state of spins plus environment is a nondegenerate singlet, as for an antiferromagnetic dimer Kondo-coupled to the substrate~\cite{jones1988}. Then exactly (SM Sec.~S6)
\begin{equation}
D_i=3,\qquad W=12-\epsilon,\qquad a=1-\epsilon/8.
\label{eq:t0}
\end{equation}
The spins are thus entangled if and only if $\epsilon<4$, that is, if $W>8$. The reduced state $\rho_{12}$ then has Werner form~\cite{werner1989} with $W_{\rm cl}=4(3-\epsilon/2)^2/(3-\epsilon/4)$. Exact diagonalization reproduces Eq.~\eqref{eq:t0} to $10^{-8}$, and NRG with continuous bands to $10^{-4}$ at $T=0.02J$ [Fig.~\ref{fig:controls}(b)].

Measured spectra are broadened by lock-in modulation and voltage noise, and they are noisy, $y_b=\bar g(V_b)+e_b$. We minimize the ratio~\eqref{eq:R} over all non-negative spectral weights and baselines that reproduce the data within tolerances $\delta_b$. We represent the weight, together with its detailed-balance partner, in narrow energy intervals. This makes the denominator of Eq.~\eqref{eq:R} exact. The broadened kernel in each interval is enclosed using its monotonicity and curvature (SM Sec.~S7). The minimization then reduces to linear programming~\cite{charnes1962,boyd2004}, in the spirit of strict-bounds approaches to inverse problems~\cite{backus1970,stark1992}. For Gaussian noise we set $\delta_b=z\sigma_b$, with $z$ chosen to cover all $n_b$ bias points simultaneously~\cite{sidak1967}. The true spectrum then satisfies the constraints with probability $1-\alpha'$, hence
\begin{equation}
\Pr\big[D_i\low\le D_i\big]\ge1-\alpha'
\label{eq:coverage}
\end{equation}
for every spectrum compatible with Eq.~\eqref{eq:forward}. The bound is also sharp. The broadened kernels determine the spectrum uniquely, so $D_i\low$ approaches Eq.~\eqref{eq:prop1} as noise and interval width vanish (SM Sec.~S7). Data incompatible with Eq.~\eqref{eq:forward} admit no solution and thus reveal violated assumptions. Local bounds on $N$ sites enter Eq.~\eqref{eq:closure} with $\alpha'=\alpha/N$.

For the isolated dimer, $H=J\bm S_1\cdot\bm S_2$, one finds with $x=J/T$
\begin{equation}
D_i=3R(x),\quad W=12R(x),\quad R(x)=\frac{(e^x-1)^2}{(e^x+3)(e^x+1)},
\label{eq:m0}
\end{equation}
and $\epsilon=0$. For exact spectra at $r=0$ the thresholds 4, 6 and 8 are exceeded for $T<0.62J$, $0.48J$ and $0.38J$ ($e^x>5$, $4+\sqrt{17}$ and $7+\sqrt{52}$). The reduced state is entangled for $T<0.91J$. At finite $r$ a threshold $W_t$ is exceeded for $r^2<\frac34(12R/W_t-1)$. Equivalently, at low temperature the bound equals $12f$, where $f$ is the fraction of the high-bias conductance carried by the spin-flip step. Thus $W>4$, 6 and 8 require $f>1/3$, $1/2$ and $2/3$. For exact data the optimized bound reproduces $D_i/(1+4r^2/3)$ to $10^{-5}$ at $T=0.05J$. With energy resolution $h=T/64$ it does so to $0.1\%$ at $0.2J$ and to $1.3\%$ at $0.5J$ [Fig.~\ref{fig:theory}(b) and SM Sec.~S8].

\begin{figure}[t]
\includegraphics[width=\columnwidth]{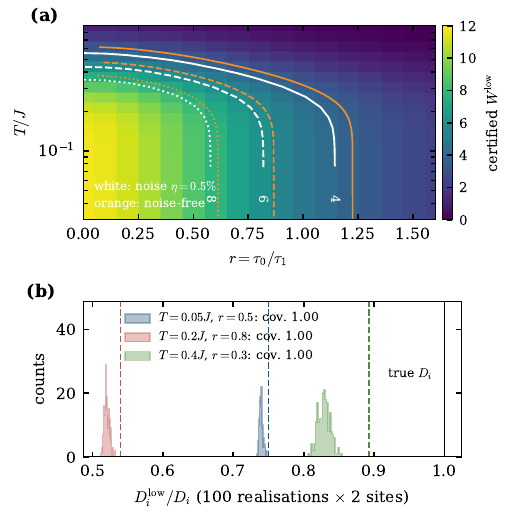}
\caption{\textbf{Certification from noisy spectra.}
(a) Bound $W\low=2(D_1\low+D_2\low)$ for the isolated dimer ($\eta=0.5\%$, 181 bias points, joint confidence level 95\%, noise-free data with tolerances). White lines are the contours $W\low=4$, 6 and 8, and orange lines are the noise-free boundaries. (b) $D_i\low/D_i$ for 100 noise realizations per site at $(T/J,r)=(0.05,0.5)$, $(0.2,0.8)$ and $(0.4,0.3)$. Dashed lines show Eq.~\eqref{eq:prop1}. The fractions exceeding 4, 6 and 8 are 1, 1 and 1, then 1, 0.48 and 0, then 1, 1 and 0.}
\label{fig:map}
\end{figure}

Figure~\ref{fig:map}(a) maps $W\low$ for noise $\eta=0.5\%$ of the high-bias conductance, 181 bias points and a joint confidence level of 95\%. The contours $W\low=4$, 6, and 8 follow the noise-free boundaries. They reach $r\approx\MapRfour$, $\MapRsix$ and $\MapReight$ at $T=0.03J$ (noise-free $\sqrt{3/2}$, $\sqrt{3/4}$ and $\sqrt{3/8}$), and $T\approx\MapTfour J$, $\MapTsix J$ and $\MapTeight J$ at $r=0$. In 100 noise realizations at each of three parameter sets, both local bounds always held [Fig.~\ref{fig:map}(b)]. Their means were $2.219\pm0.008$, $1.500\pm0.011$ and $1.554\pm0.016$, compared with 2.250, 1.555, and 1.674 from Eq.~\eqref{eq:prop1}. We also tested 945 settings of $T$, $r$, noise level, lock-in amplitude and number of bias points, with 2000 realizations each. The joint coverage never fell below $0.999$. With $10\%$ errors in the assumed instrumental parameters, it stayed at or above $0.996$. In the reference configuration $W\low$ exceeds 8, 6 and 4 in at least $99\%$ of realizations for $r\le0.3$, $0.6$ and $0.9$, respectively, at $T\le0.3J$ (SM Sec.~S8).

Figure~\ref{fig:controls}(a) contrasts a naive analysis that ignores leakage with the full-state $W$ and the reduced $W_{\rm cl}$. A ferromagnetic dimer gives $W=4$ and is never certified. A N\'eel product state appears as $W=8$ in the naive analysis, but its leakage $\epsilon=4$ restores $W=4$. A spin-1 ion misidentified as a spin-$\frac12$ dimer would appear as $W=12$. The probed sites must therefore be established independently as distinct spin-$\frac12$ moments, for example, by spatial mapping. Two spins, each screened by a private partner, produce local spectra identical to those of the entangled dimer. Their $W=6$ exceeds the product bound while $\rho_{12}$ is maximally mixed. The criteria $W>6$ and $W>8$ correctly reject this false positive. Cluster identification is thus an assumption that no spectroscopic method can dispense with. Given it, $W>6$ and $W>8$ close the environment loophole. The bath-mediated state is correctly certified by $W>6$ and rejected by $W>8$. For the Kondo-coupled dimer with independent bands [Fig.~\ref{fig:controls}(b)], $W>6$ holds deep into the screened regime ($\epsilon<6$ at $T=0$), whereas $W>8$ fails exactly where $\rho_{12}$ becomes separable. In the NRG, Kondo-screened dimers have $\epsilon=5.0$--5.9, $W_{\rm cl}\le0.55$ and $W=6.1$--6.9 at $T=0.02J$. With a shared band, strong coupling can instead leave a degenerate, underscreened ground state. Then $W$ drops below 6, for example to $W=4.6$ for $S=0.6$ and $\nu J_K=0.2$, and no threshold is passed. At finite temperature $W-W_{\rm cl}$ stayed below $2\epsilon$ in all 360 discretized-bath configurations examined, up to $J_K=4J$, in all 90 NRG points with independent bands, and at all reliable NRG points with a shared band. This is, however, not guaranteed in general (SM Sec.~S6).

The certificate relies on a small set of experimentally testable conditions. The probed object must be a correctly identified spin cluster with well-defined local moments, and the tunneling must remain weak enough that the spins stay in thermal equilibrium (SM Sec.~S1). The spin temperature and instrumental broadening must also be known within specified bounds. These assumptions can be checked directly against the measured spectra. When they are violated, the data either become incompatible with the forward model in Eq.~\eqref{eq:forward} or lead to a controlled weakening of the bound. Charge fluctuations provide one example. If the double occupancy is $d$, the inferred $D_i\low$ can be overestimated by at most a factor $(1-2d)^{-1}$. For a Hubbard dimer with $U/t=16$, $d=1.5\%$, so the correction is only a few percent (SM Sec.~S8).

The experimental requirements are within reach of present atomic-scale spectroscopy. The certificate requires resolved spin-excitation steps at $T\lesssim0.3$--$0.5J$, with sub-percent conductance noise over a few hundred bias points. These conditions are already approached in transition-metal atoms and dimers on MgO/Ag(001)~\cite{baumann2015,yang2019,wang2023}, magnetic atoms on Cu$_2$N~\cite{hirjibehedin2006}, and nanographene spin dimers and chains on Au(111)~\cite{mishra2021}. Unlike ESR-STM protocols~\cite{delcastillo2023,broekhoven2024}, the present method requires no coherent driving. Unlike susceptibility-based QFI witnesses~\cite{hauke2016,scheie2021}, it requires no absolute intensity calibration. A natural first experiment is an exchange-coupled dimer, where the excitation energy, linewidth, and local spectra can all be measured independently. Temperature-dependent measurements could then follow the loss of entanglement as thermal fluctuations or Kondo screening compete with interspin exchange.

More broadly, the work shows how a nonlocal quantum property can be certified from local equilibrium measurements when exact sum rules fix the available fluctuation weight and conservation laws constrain how that weight is shared. For larger spin structures, the same framework identifies the additional intersite information needed once local spectra alone are no longer sufficient. Combined with spatially resolved spectroscopy or ESR-STM, this could map quantum entanglement across engineered spin networks and test it before applying coherent operations. Local tunneling spectroscopy can therefore provide not only excitation energies and coupling strengths, but also a quantitative and statistically controlled measure of the quantum correlations carried by individual spins.

The data that support the findings of this article are available from the authors upon reasonable request.

\begin{acknowledgments}
Funding information: [to be added].
\end{acknowledgments}

\bibliography{references}

\begin{thebibliography}{47}%
\makeatletter
\providecommand \@ifxundefined [1]{%
 \@ifx{#1\undefined}
}%
\providecommand \@ifnum [1]{%
 \ifnum #1\expandafter \@firstoftwo
 \else \expandafter \@secondoftwo
 \fi
}%
\providecommand \@ifx [1]{%
 \ifx #1\expandafter \@firstoftwo
 \else \expandafter \@secondoftwo
 \fi
}%
\providecommand \natexlab [1]{#1}%
\providecommand \enquote  [1]{``#1''}%
\providecommand \bibnamefont  [1]{#1}%
\providecommand \bibfnamefont [1]{#1}%
\providecommand \citenamefont [1]{#1}%
\providecommand \href@noop [0]{\@secondoftwo}%
\providecommand \href [0]{\begingroup \@sanitize@url \@href}%
\providecommand \@href[1]{\@@startlink{#1}\@@href}%
\providecommand \@@href[1]{\endgroup#1\@@endlink}%
\providecommand \@sanitize@url [0]{\catcode `\\12\catcode `\$12\catcode
  `\&12\catcode `\#12\catcode `\^12\catcode `\_12\catcode `\%12\relax}%
\providecommand \@@startlink[1]{}%
\providecommand \@@endlink[0]{}%
\providecommand \url  [0]{\begingroup\@sanitize@url \@url }%
\providecommand \@url [1]{\endgroup\@href {#1}{\urlprefix }}%
\providecommand \urlprefix  [0]{URL }%
\providecommand \Eprint [0]{\href }%
\providecommand \doibase [0]{https://doi.org/}%
\providecommand \selectlanguage [0]{\@gobble}%
\providecommand \bibinfo  [0]{\@secondoftwo}%
\providecommand \bibfield  [0]{\@secondoftwo}%
\providecommand \translation [1]{[#1]}%
\providecommand \BibitemOpen [0]{}%
\providecommand \bibitemStop [0]{}%
\providecommand \bibitemNoStop [0]{.\EOS\space}%
\providecommand \EOS [0]{\spacefactor3000\relax}%
\providecommand \BibitemShut  [1]{\csname bibitem#1\endcsname}%
\let\auto@bib@innerbib\@empty
\bibitem [{\citenamefont {G\"uhne}\ and\ \citenamefont
  {T\'oth}(2009)}]{guhne2009}%
  \BibitemOpen
  \bibfield  {author} {\bibinfo {author} {\bibfnamefont {O.}~\bibnamefont
  {G\"uhne}}\ and\ \bibinfo {author} {\bibfnamefont {G.}~\bibnamefont
  {T\'oth}},\ }\bibfield  {title} {\bibinfo {title} {Entanglement detection},\
  }\href@noop {} {\bibfield  {journal} {\bibinfo  {journal} {Phys. Rep.}\
  }\textbf {\bibinfo {volume} {474}},\ \bibinfo {pages} {1} (\bibinfo {year}
  {2009})}\BibitemShut {NoStop}%
\bibitem [{\citenamefont {Pezz\`e}\ \emph {et~al.}(2018)\citenamefont
  {Pezz\`e}, \citenamefont {Smerzi}, \citenamefont {Oberthaler}, \citenamefont
  {Schmied},\ and\ \citenamefont {Treutlein}}]{pezze2018}%
  \BibitemOpen
  \bibfield  {author} {\bibinfo {author} {\bibfnamefont {L.}~\bibnamefont
  {Pezz\`e}}, \bibinfo {author} {\bibfnamefont {A.}~\bibnamefont {Smerzi}},
  \bibinfo {author} {\bibfnamefont {M.~K.}\ \bibnamefont {Oberthaler}},
  \bibinfo {author} {\bibfnamefont {R.}~\bibnamefont {Schmied}},\ and\ \bibinfo
  {author} {\bibfnamefont {P.}~\bibnamefont {Treutlein}},\ }\bibfield  {title}
  {\bibinfo {title} {Quantum metrology with nonclassical states of atomic
  ensembles},\ }\href {https://doi.org/10.1103/RevModPhys.90.035005} {\bibfield
   {journal} {\bibinfo  {journal} {Rev. Mod. Phys.}\ }\textbf {\bibinfo
  {volume} {90}},\ \bibinfo {pages} {035005} (\bibinfo {year}
  {2018})}\BibitemShut {NoStop}%
\bibitem [{\citenamefont {Braunstein}\ and\ \citenamefont
  {Caves}(1994)}]{braunstein1994}%
  \BibitemOpen
  \bibfield  {author} {\bibinfo {author} {\bibfnamefont {S.~L.}\ \bibnamefont
  {Braunstein}}\ and\ \bibinfo {author} {\bibfnamefont {C.~M.}\ \bibnamefont
  {Caves}},\ }\bibfield  {title} {\bibinfo {title} {Statistical distance and
  the geometry of quantum states},\ }\href
  {https://doi.org/10.1103/PhysRevLett.72.3439} {\bibfield  {journal} {\bibinfo
   {journal} {Phys. Rev. Lett.}\ }\textbf {\bibinfo {volume} {72}},\ \bibinfo
  {pages} {3439} (\bibinfo {year} {1994})}\BibitemShut {NoStop}%
\bibitem [{\citenamefont {Pezz\`e}\ and\ \citenamefont
  {Smerzi}(2009)}]{pezze2009}%
  \BibitemOpen
  \bibfield  {author} {\bibinfo {author} {\bibfnamefont {L.}~\bibnamefont
  {Pezz\`e}}\ and\ \bibinfo {author} {\bibfnamefont {A.}~\bibnamefont
  {Smerzi}},\ }\bibfield  {title} {\bibinfo {title} {Entanglement, nonlinear
  dynamics, and the {H}eisenberg limit},\ }\href
  {https://doi.org/10.1103/PhysRevLett.102.100401} {\bibfield  {journal}
  {\bibinfo  {journal} {Phys. Rev. Lett.}\ }\textbf {\bibinfo {volume} {102}},\
  \bibinfo {pages} {100401} (\bibinfo {year} {2009})}\BibitemShut {NoStop}%
\bibitem [{\citenamefont {Hyllus}\ \emph {et~al.}(2012)\citenamefont {Hyllus},
  \citenamefont {Laskowski}, \citenamefont {Krischek}, \citenamefont
  {Schwemmer}, \citenamefont {Wieczorek}, \citenamefont {Weinfurter},
  \citenamefont {Pezz\`e},\ and\ \citenamefont {Smerzi}}]{hyllus2012}%
  \BibitemOpen
  \bibfield  {author} {\bibinfo {author} {\bibfnamefont {P.}~\bibnamefont
  {Hyllus}}, \bibinfo {author} {\bibfnamefont {W.}~\bibnamefont {Laskowski}},
  \bibinfo {author} {\bibfnamefont {R.}~\bibnamefont {Krischek}}, \bibinfo
  {author} {\bibfnamefont {C.}~\bibnamefont {Schwemmer}}, \bibinfo {author}
  {\bibfnamefont {W.}~\bibnamefont {Wieczorek}}, \bibinfo {author}
  {\bibfnamefont {H.}~\bibnamefont {Weinfurter}}, \bibinfo {author}
  {\bibfnamefont {L.}~\bibnamefont {Pezz\`e}},\ and\ \bibinfo {author}
  {\bibfnamefont {A.}~\bibnamefont {Smerzi}},\ }\bibfield  {title} {\bibinfo
  {title} {Fisher information and multiparticle entanglement},\ }\href
  {https://doi.org/10.1103/PhysRevA.85.022321} {\bibfield  {journal} {\bibinfo
  {journal} {Phys. Rev. A}\ }\textbf {\bibinfo {volume} {85}},\ \bibinfo
  {pages} {022321} (\bibinfo {year} {2012})}\BibitemShut {NoStop}%
\bibitem [{\citenamefont {T\'oth}(2012)}]{toth2012}%
  \BibitemOpen
  \bibfield  {author} {\bibinfo {author} {\bibfnamefont {G.}~\bibnamefont
  {T\'oth}},\ }\bibfield  {title} {\bibinfo {title} {Multipartite entanglement
  and high-precision metrology},\ }\href
  {https://doi.org/10.1103/PhysRevA.85.022322} {\bibfield  {journal} {\bibinfo
  {journal} {Phys. Rev. A}\ }\textbf {\bibinfo {volume} {85}},\ \bibinfo
  {pages} {022322} (\bibinfo {year} {2012})}\BibitemShut {NoStop}%
\bibitem [{\citenamefont {Hauke}\ \emph {et~al.}(2016)\citenamefont {Hauke},
  \citenamefont {Heyl}, \citenamefont {Tagliacozzo},\ and\ \citenamefont
  {Zoller}}]{hauke2016}%
  \BibitemOpen
  \bibfield  {author} {\bibinfo {author} {\bibfnamefont {P.}~\bibnamefont
  {Hauke}}, \bibinfo {author} {\bibfnamefont {M.}~\bibnamefont {Heyl}},
  \bibinfo {author} {\bibfnamefont {L.}~\bibnamefont {Tagliacozzo}},\ and\
  \bibinfo {author} {\bibfnamefont {P.}~\bibnamefont {Zoller}},\ }\bibfield
  {title} {\bibinfo {title} {Measuring multipartite entanglement through
  dynamic susceptibilities},\ }\href {https://doi.org/10.1038/nphys3700}
  {\bibfield  {journal} {\bibinfo  {journal} {Nat. Phys.}\ }\textbf {\bibinfo
  {volume} {12}},\ \bibinfo {pages} {778} (\bibinfo {year} {2016})}\BibitemShut
  {NoStop}%
\bibitem [{\citenamefont {Shitara}\ and\ \citenamefont
  {Ueda}(2016)}]{shitara2016}%
  \BibitemOpen
  \bibfield  {author} {\bibinfo {author} {\bibfnamefont {T.}~\bibnamefont
  {Shitara}}\ and\ \bibinfo {author} {\bibfnamefont {M.}~\bibnamefont {Ueda}},\
  }\bibfield  {title} {\bibinfo {title} {Determining the continuous family of
  quantum {F}isher information from linear-response theory},\ }\href
  {https://doi.org/10.1103/PhysRevA.94.062316} {\bibfield  {journal} {\bibinfo
  {journal} {Phys. Rev. A}\ }\textbf {\bibinfo {volume} {94}},\ \bibinfo
  {pages} {062316} (\bibinfo {year} {2016})}\BibitemShut {NoStop}%
\bibitem [{\citenamefont {Mathew}\ \emph {et~al.}(2020)\citenamefont {Mathew}
  \emph {et~al.}}]{mathew2020}%
  \BibitemOpen
  \bibfield  {author} {\bibinfo {author} {\bibfnamefont {G.}~\bibnamefont
  {Mathew}} \emph {et~al.},\ }\bibfield  {title} {\bibinfo {title}
  {Experimental realization of multipartite entanglement via quantum {F}isher
  information in a uniform antiferromagnetic quantum spin chain},\ }\href
  {https://doi.org/10.1103/PhysRevResearch.2.043329} {\bibfield  {journal}
  {\bibinfo  {journal} {Phys. Rev. Research}\ }\textbf {\bibinfo {volume}
  {2}},\ \bibinfo {pages} {043329} (\bibinfo {year} {2020})}\BibitemShut
  {NoStop}%
\bibitem [{\citenamefont {Scheie}\ \emph {et~al.}(2021)\citenamefont {Scheie},
  \citenamefont {Laurell}, \citenamefont {Samarakoon} \emph
  {et~al.}}]{scheie2021}%
  \BibitemOpen
  \bibfield  {author} {\bibinfo {author} {\bibfnamefont {A.}~\bibnamefont
  {Scheie}}, \bibinfo {author} {\bibfnamefont {P.}~\bibnamefont {Laurell}},
  \bibinfo {author} {\bibfnamefont {A.~M.}\ \bibnamefont {Samarakoon}}, \emph
  {et~al.},\ }\bibfield  {title} {\bibinfo {title} {Witnessing entanglement in
  quantum magnets using neutron scattering},\ }\href
  {https://doi.org/10.1103/PhysRevB.103.224434} {\bibfield  {journal} {\bibinfo
   {journal} {Phys. Rev. B}\ }\textbf {\bibinfo {volume} {103}},\ \bibinfo
  {pages} {224434} (\bibinfo {year} {2021})}\BibitemShut {NoStop}%
\bibitem [{\citenamefont {Laurell}\ \emph {et~al.}(2021)\citenamefont
  {Laurell}, \citenamefont {Scheie}, \citenamefont {Mukherjee} \emph
  {et~al.}}]{laurell2021}%
  \BibitemOpen
  \bibfield  {author} {\bibinfo {author} {\bibfnamefont {P.}~\bibnamefont
  {Laurell}}, \bibinfo {author} {\bibfnamefont {A.}~\bibnamefont {Scheie}},
  \bibinfo {author} {\bibfnamefont {C.~J.}\ \bibnamefont {Mukherjee}}, \emph
  {et~al.},\ }\bibfield  {title} {\bibinfo {title} {Quantifying and controlling
  entanglement in the quantum magnet {C}s$_2${C}o{C}l$_4$},\ }\href
  {https://doi.org/10.1103/PhysRevLett.127.037201} {\bibfield  {journal}
  {\bibinfo  {journal} {Phys. Rev. Lett.}\ }\textbf {\bibinfo {volume} {127}},\
  \bibinfo {pages} {037201} (\bibinfo {year} {2021})}\BibitemShut {NoStop}%
\bibitem [{\citenamefont {Yin}\ \emph {et~al.}(2021)\citenamefont {Yin},
  \citenamefont {Pan},\ and\ \citenamefont {Hasan}}]{yin2021}%
  \BibitemOpen
  \bibfield  {author} {\bibinfo {author} {\bibfnamefont {J.-X.}\ \bibnamefont
  {Yin}}, \bibinfo {author} {\bibfnamefont {S.~H.}\ \bibnamefont {Pan}},\ and\
  \bibinfo {author} {\bibfnamefont {M.~Z.}\ \bibnamefont {Hasan}},\ }\bibfield
  {title} {\bibinfo {title} {Probing topological quantum matter with scanning
  tunnelling microscopy},\ }\href {https://doi.org/10.1038/s42254-021-00293-7}
  {\bibfield  {journal} {\bibinfo  {journal} {Nat. Rev. Phys.}\ }\textbf
  {\bibinfo {volume} {3}},\ \bibinfo {pages} {249} (\bibinfo {year}
  {2021})}\BibitemShut {NoStop}%
\bibitem [{\citenamefont {Heinrich}\ \emph {et~al.}(2004)\citenamefont
  {Heinrich}, \citenamefont {Gupta}, \citenamefont {Lutz},\ and\ \citenamefont
  {Eigler}}]{heinrich2004}%
  \BibitemOpen
  \bibfield  {author} {\bibinfo {author} {\bibfnamefont {A.~J.}\ \bibnamefont
  {Heinrich}}, \bibinfo {author} {\bibfnamefont {J.~A.}\ \bibnamefont {Gupta}},
  \bibinfo {author} {\bibfnamefont {C.~P.}\ \bibnamefont {Lutz}},\ and\
  \bibinfo {author} {\bibfnamefont {D.~M.}\ \bibnamefont {Eigler}},\ }\bibfield
   {title} {\bibinfo {title} {Single-atom spin-flip spectroscopy},\ }\href
  {https://doi.org/10.1126/science.1101077} {\bibfield  {journal} {\bibinfo
  {journal} {Science}\ }\textbf {\bibinfo {volume} {306}},\ \bibinfo {pages}
  {466} (\bibinfo {year} {2004})}\BibitemShut {NoStop}%
\bibitem [{\citenamefont {Hirjibehedin}\ \emph {et~al.}(2006)\citenamefont
  {Hirjibehedin}, \citenamefont {Lutz},\ and\ \citenamefont
  {Heinrich}}]{hirjibehedin2006}%
  \BibitemOpen
  \bibfield  {author} {\bibinfo {author} {\bibfnamefont {C.~F.}\ \bibnamefont
  {Hirjibehedin}}, \bibinfo {author} {\bibfnamefont {C.~P.}\ \bibnamefont
  {Lutz}},\ and\ \bibinfo {author} {\bibfnamefont {A.~J.}\ \bibnamefont
  {Heinrich}},\ }\bibfield  {title} {\bibinfo {title} {Spin coupling in
  engineered atomic structures},\ }\href@noop {} {\bibfield  {journal}
  {\bibinfo  {journal} {Science}\ }\textbf {\bibinfo {volume} {312}},\ \bibinfo
  {pages} {1021} (\bibinfo {year} {2006})}\BibitemShut {NoStop}%
\bibitem [{\citenamefont {Fern\'andez-Rossier}(2009)}]{fernandezrossier2009}%
  \BibitemOpen
  \bibfield  {author} {\bibinfo {author} {\bibfnamefont {J.}~\bibnamefont
  {Fern\'andez-Rossier}},\ }\bibfield  {title} {\bibinfo {title} {Theory of
  single-spin inelastic tunneling spectroscopy},\ }\href
  {https://doi.org/10.1103/PhysRevLett.102.256802} {\bibfield  {journal}
  {\bibinfo  {journal} {Phys. Rev. Lett.}\ }\textbf {\bibinfo {volume} {102}},\
  \bibinfo {pages} {256802} (\bibinfo {year} {2009})}\BibitemShut {NoStop}%
\bibitem [{\citenamefont {Lorente}\ and\ \citenamefont
  {Gauyacq}(2009)}]{lorente2009}%
  \BibitemOpen
  \bibfield  {author} {\bibinfo {author} {\bibfnamefont {N.}~\bibnamefont
  {Lorente}}\ and\ \bibinfo {author} {\bibfnamefont {J.-P.}\ \bibnamefont
  {Gauyacq}},\ }\bibfield  {title} {\bibinfo {title} {Efficient spin
  transitions in inelastic electron tunneling spectroscopy},\ }\href
  {https://doi.org/10.1103/PhysRevLett.103.176601} {\bibfield  {journal}
  {\bibinfo  {journal} {Phys. Rev. Lett.}\ }\textbf {\bibinfo {volume} {103}},\
  \bibinfo {pages} {176601} (\bibinfo {year} {2009})}\BibitemShut {NoStop}%
\bibitem [{\citenamefont {Ternes}(2015)}]{ternes2015}%
  \BibitemOpen
  \bibfield  {author} {\bibinfo {author} {\bibfnamefont {M.}~\bibnamefont
  {Ternes}},\ }\bibfield  {title} {\bibinfo {title} {Spin excitations and
  correlations in scanning tunneling spectroscopy},\ }\href
  {https://doi.org/10.1088/1367-2630/17/6/063016} {\bibfield  {journal}
  {\bibinfo  {journal} {New J. Phys.}\ }\textbf {\bibinfo {volume} {17}},\
  \bibinfo {pages} {063016} (\bibinfo {year} {2015})}\BibitemShut {NoStop}%
\bibitem [{\citenamefont {Ternes}(2017)}]{ternes2017}%
  \BibitemOpen
  \bibfield  {author} {\bibinfo {author} {\bibfnamefont {M.}~\bibnamefont
  {Ternes}},\ }\bibfield  {title} {\bibinfo {title} {Probing magnetic
  excitations and correlations in single and coupled spin systems with scanning
  tunneling spectroscopy},\ }\href@noop {} {\bibfield  {journal} {\bibinfo
  {journal} {Prog. Surf. Sci.}\ }\textbf {\bibinfo {volume} {92}},\ \bibinfo
  {pages} {83} (\bibinfo {year} {2017})}\BibitemShut {NoStop}%
\bibitem [{\citenamefont {Choi}\ \emph {et~al.}(2019)\citenamefont {Choi},
  \citenamefont {Lorente}, \citenamefont {Wiebe}, \citenamefont {von Bergmann},
  \citenamefont {Otte},\ and\ \citenamefont {Heinrich}}]{choi2019}%
  \BibitemOpen
  \bibfield  {author} {\bibinfo {author} {\bibfnamefont {D.-J.}\ \bibnamefont
  {Choi}}, \bibinfo {author} {\bibfnamefont {N.}~\bibnamefont {Lorente}},
  \bibinfo {author} {\bibfnamefont {J.}~\bibnamefont {Wiebe}}, \bibinfo
  {author} {\bibfnamefont {K.}~\bibnamefont {von Bergmann}}, \bibinfo {author}
  {\bibfnamefont {A.~F.}\ \bibnamefont {Otte}},\ and\ \bibinfo {author}
  {\bibfnamefont {A.~J.}\ \bibnamefont {Heinrich}},\ }\bibfield  {title}
  {\bibinfo {title} {Colloquium: Atomic spin chains on surfaces},\ }\href
  {https://doi.org/10.1103/RevModPhys.91.041001} {\bibfield  {journal}
  {\bibinfo  {journal} {Rev. Mod. Phys.}\ }\textbf {\bibinfo {volume} {91}},\
  \bibinfo {pages} {041001} (\bibinfo {year} {2019})}\BibitemShut {NoStop}%
\bibitem [{\citenamefont {Mishra}\ \emph {et~al.}(2021)\citenamefont {Mishra},
  \citenamefont {Catarina}, \citenamefont {Wu} \emph {et~al.}}]{mishra2021}%
  \BibitemOpen
  \bibfield  {author} {\bibinfo {author} {\bibfnamefont {S.}~\bibnamefont
  {Mishra}}, \bibinfo {author} {\bibfnamefont {G.}~\bibnamefont {Catarina}},
  \bibinfo {author} {\bibfnamefont {F.}~\bibnamefont {Wu}}, \emph {et~al.},\
  }\bibfield  {title} {\bibinfo {title} {Observation of fractional edge
  excitations in nanographene spin chains},\ }\href
  {https://doi.org/10.1038/s41586-021-03842-3} {\bibfield  {journal} {\bibinfo
  {journal} {Nature}\ }\textbf {\bibinfo {volume} {598}},\ \bibinfo {pages}
  {287} (\bibinfo {year} {2021})}\BibitemShut {NoStop}%
\bibitem [{\citenamefont {Krammer}\ \emph {et~al.}(2009)\citenamefont
  {Krammer}, \citenamefont {Kampermann}, \citenamefont {Bru\ss{}},
  \citenamefont {Bertlmann}, \citenamefont {Kwek},\ and\ \citenamefont
  {Macchiavello}}]{krammer2009}%
  \BibitemOpen
  \bibfield  {author} {\bibinfo {author} {\bibfnamefont {P.}~\bibnamefont
  {Krammer}}, \bibinfo {author} {\bibfnamefont {H.}~\bibnamefont {Kampermann}},
  \bibinfo {author} {\bibfnamefont {D.}~\bibnamefont {Bru\ss{}}}, \bibinfo
  {author} {\bibfnamefont {R.~A.}\ \bibnamefont {Bertlmann}}, \bibinfo {author}
  {\bibfnamefont {L.~C.}\ \bibnamefont {Kwek}},\ and\ \bibinfo {author}
  {\bibfnamefont {C.}~\bibnamefont {Macchiavello}},\ }\bibfield  {title}
  {\bibinfo {title} {Multipartite entanglement detection via structure
  factors},\ }\href {https://doi.org/10.1103/PhysRevLett.103.100502} {\bibfield
   {journal} {\bibinfo  {journal} {Phys. Rev. Lett.}\ }\textbf {\bibinfo
  {volume} {103}},\ \bibinfo {pages} {100502} (\bibinfo {year}
  {2009})}\BibitemShut {NoStop}%
\bibitem [{\citenamefont {Wie\'sniak}\ \emph {et~al.}(2005)\citenamefont
  {Wie\'sniak}, \citenamefont {Vedral},\ and\ \citenamefont {\v{C}.
  Brukner}}]{wiesniak2005}%
  \BibitemOpen
  \bibfield  {author} {\bibinfo {author} {\bibfnamefont {M.}~\bibnamefont
  {Wie\'sniak}}, \bibinfo {author} {\bibfnamefont {V.}~\bibnamefont {Vedral}},\
  and\ \bibinfo {author} {\bibnamefont {\v{C}. Brukner}},\ }\bibfield  {title}
  {\bibinfo {title} {Magnetic susceptibility as a macroscopic entanglement
  witness},\ }\href {https://doi.org/10.1088/1367-2630/7/1/258} {\bibfield
  {journal} {\bibinfo  {journal} {New J. Phys.}\ }\textbf {\bibinfo {volume}
  {7}},\ \bibinfo {pages} {258} (\bibinfo {year} {2005})}\BibitemShut {NoStop}%
\bibitem [{\citenamefont {Malla}\ \emph {et~al.}(2024)\citenamefont {Malla},
  \citenamefont {Weichselbaum}, \citenamefont {Wei},\ and\ \citenamefont
  {Konik}}]{malla2024}%
  \BibitemOpen
  \bibfield  {author} {\bibinfo {author} {\bibfnamefont {R.~K.}\ \bibnamefont
  {Malla}}, \bibinfo {author} {\bibfnamefont {A.}~\bibnamefont {Weichselbaum}},
  \bibinfo {author} {\bibfnamefont {T.-C.}\ \bibnamefont {Wei}},\ and\ \bibinfo
  {author} {\bibfnamefont {R.~M.}\ \bibnamefont {Konik}},\ }\bibfield  {title}
  {\bibinfo {title} {Detecting multipartite entanglement patterns using
  single-particle {G}reen's functions},\ }\href
  {https://doi.org/10.1103/PhysRevLett.133.260202} {\bibfield  {journal}
  {\bibinfo  {journal} {Phys. Rev. Lett.}\ }\textbf {\bibinfo {volume} {133}},\
  \bibinfo {pages} {260202} (\bibinfo {year} {2024})}\BibitemShut {NoStop}%
\bibitem [{\citenamefont {Baumann}\ \emph {et~al.}(2015)\citenamefont
  {Baumann}, \citenamefont {Paul}, \citenamefont {Choi}, \citenamefont {Lutz},
  \citenamefont {Ardavan},\ and\ \citenamefont {Heinrich}}]{baumann2015}%
  \BibitemOpen
  \bibfield  {author} {\bibinfo {author} {\bibfnamefont {S.}~\bibnamefont
  {Baumann}}, \bibinfo {author} {\bibfnamefont {W.}~\bibnamefont {Paul}},
  \bibinfo {author} {\bibfnamefont {T.}~\bibnamefont {Choi}}, \bibinfo {author}
  {\bibfnamefont {C.~P.}\ \bibnamefont {Lutz}}, \bibinfo {author}
  {\bibfnamefont {A.}~\bibnamefont {Ardavan}},\ and\ \bibinfo {author}
  {\bibfnamefont {A.~J.}\ \bibnamefont {Heinrich}},\ }\bibfield  {title}
  {\bibinfo {title} {Electron paramagnetic resonance of individual atoms on a
  surface},\ }\href {https://doi.org/10.1126/science.aac8703} {\bibfield
  {journal} {\bibinfo  {journal} {Science}\ }\textbf {\bibinfo {volume}
  {350}},\ \bibinfo {pages} {417} (\bibinfo {year} {2015})}\BibitemShut
  {NoStop}%
\bibitem [{\citenamefont {Yang}\ \emph {et~al.}(2019)\citenamefont {Yang},
  \citenamefont {Paul}, \citenamefont {Phark} \emph {et~al.}}]{yang2019}%
  \BibitemOpen
  \bibfield  {author} {\bibinfo {author} {\bibfnamefont {K.}~\bibnamefont
  {Yang}}, \bibinfo {author} {\bibfnamefont {W.}~\bibnamefont {Paul}}, \bibinfo
  {author} {\bibfnamefont {S.-H.}\ \bibnamefont {Phark}}, \emph {et~al.},\
  }\bibfield  {title} {\bibinfo {title} {Coherent spin manipulation of
  individual atoms on a surface},\ }\href
  {https://doi.org/10.1126/science.aay6779} {\bibfield  {journal} {\bibinfo
  {journal} {Science}\ }\textbf {\bibinfo {volume} {366}},\ \bibinfo {pages}
  {509} (\bibinfo {year} {2019})}\BibitemShut {NoStop}%
\bibitem [{\citenamefont {Wang}\ \emph {et~al.}(2023)\citenamefont {Wang} \emph
  {et~al.}}]{wang2023}%
  \BibitemOpen
  \bibfield  {author} {\bibinfo {author} {\bibfnamefont {Y.}~\bibnamefont
  {Wang}} \emph {et~al.},\ }\bibfield  {title} {\bibinfo {title} {An
  atomic-scale multi-qubit platform},\ }\href
  {https://doi.org/10.1126/science.ade5050} {\bibfield  {journal} {\bibinfo
  {journal} {Science}\ }\textbf {\bibinfo {volume} {382}},\ \bibinfo {pages}
  {87} (\bibinfo {year} {2023})}\BibitemShut {NoStop}%
\bibitem [{\citenamefont {del Castillo}\ and\ \citenamefont
  {Fern\'andez-Rossier}(2023)}]{delcastillo2023}%
  \BibitemOpen
  \bibfield  {author} {\bibinfo {author} {\bibfnamefont {Y.}~\bibnamefont {del
  Castillo}}\ and\ \bibinfo {author} {\bibfnamefont {J.}~\bibnamefont
  {Fern\'andez-Rossier}},\ }\bibfield  {title} {\bibinfo {title} {Certifying
  entanglement of spins on surfaces using {ESR-STM}},\ }\href
  {https://doi.org/10.1103/PhysRevB.108.115413} {\bibfield  {journal} {\bibinfo
   {journal} {Phys. Rev. B}\ }\textbf {\bibinfo {volume} {108}},\ \bibinfo
  {pages} {115413} (\bibinfo {year} {2023})}\BibitemShut {NoStop}%
\bibitem [{\citenamefont {Broekhoven}\ \emph {et~al.}(2024)\citenamefont
  {Broekhoven}, \citenamefont {Lee}, \citenamefont {h.~Phark}, \citenamefont
  {Otte},\ and\ \citenamefont {Wolf}}]{broekhoven2024}%
  \BibitemOpen
  \bibfield  {author} {\bibinfo {author} {\bibfnamefont {R.}~\bibnamefont
  {Broekhoven}}, \bibinfo {author} {\bibfnamefont {C.}~\bibnamefont {Lee}},
  \bibinfo {author} {\bibfnamefont {S.}~\bibnamefont {h.~Phark}}, \bibinfo
  {author} {\bibfnamefont {S.}~\bibnamefont {Otte}},\ and\ \bibinfo {author}
  {\bibfnamefont {C.}~\bibnamefont {Wolf}},\ }\bibfield  {title} {\bibinfo
  {title} {Protocol for certifying entanglement in surface spin systems using a
  scanning tunneling microscope},\ }\href
  {https://doi.org/10.1038/s41534-024-00888-9} {\bibfield  {journal} {\bibinfo
  {journal} {npj Quantum Inf.}\ }\textbf {\bibinfo {volume} {10}},\ \bibinfo
  {pages} {92} (\bibinfo {year} {2024})}\BibitemShut {NoStop}%
\bibitem [{\citenamefont {Appelbaum}(1967)}]{appelbaum1967}%
  \BibitemOpen
  \bibfield  {author} {\bibinfo {author} {\bibfnamefont {J.~A.}\ \bibnamefont
  {Appelbaum}},\ }\bibfield  {title} {\bibinfo {title} {Exchange model of
  zero-bias tunneling anomalies},\ }\href
  {https://doi.org/10.1103/PhysRev.154.633} {\bibfield  {journal} {\bibinfo
  {journal} {Phys. Rev.}\ }\textbf {\bibinfo {volume} {154}},\ \bibinfo {pages}
  {633} (\bibinfo {year} {1967})}\BibitemShut {NoStop}%
\bibitem [{SM()}]{SM}%
  \BibitemOpen
  \href@noop {} {}\bibinfo {note} {See the Supplemental Material appended to
  this article for derivations, details of the tunneling model and of the
  statistical analysis of noisy spectra, and additional numerical
  results}\BibitemShut {NoStop}%
\bibitem [{\citenamefont {Loth}\ \emph {et~al.}(2010)\citenamefont {Loth},
  \citenamefont {Etzkorn}, \citenamefont {Lutz}, \citenamefont {Eigler},\ and\
  \citenamefont {Heinrich}}]{loth2010}%
  \BibitemOpen
  \bibfield  {author} {\bibinfo {author} {\bibfnamefont {S.}~\bibnamefont
  {Loth}}, \bibinfo {author} {\bibfnamefont {M.}~\bibnamefont {Etzkorn}},
  \bibinfo {author} {\bibfnamefont {C.~P.}\ \bibnamefont {Lutz}}, \bibinfo
  {author} {\bibfnamefont {D.~M.}\ \bibnamefont {Eigler}},\ and\ \bibinfo
  {author} {\bibfnamefont {A.~J.}\ \bibnamefont {Heinrich}},\ }\bibfield
  {title} {\bibinfo {title} {Measurement of fast electron spin relaxation times
  with atomic resolution},\ }\href {https://doi.org/10.1126/science.1191688}
  {\bibfield  {journal} {\bibinfo  {journal} {Science}\ }\textbf {\bibinfo
  {volume} {329}},\ \bibinfo {pages} {1628} (\bibinfo {year}
  {2010})}\BibitemShut {NoStop}%
\bibitem [{\citenamefont {Wilson}(1975)}]{wilson1975}%
  \BibitemOpen
  \bibfield  {author} {\bibinfo {author} {\bibfnamefont {K.~G.}\ \bibnamefont
  {Wilson}},\ }\bibfield  {title} {\bibinfo {title} {The renormalization group:
  Critical phenomena and the {K}ondo problem},\ }\href
  {https://doi.org/10.1103/RevModPhys.47.773} {\bibfield  {journal} {\bibinfo
  {journal} {Rev. Mod. Phys.}\ }\textbf {\bibinfo {volume} {47}},\ \bibinfo
  {pages} {773} (\bibinfo {year} {1975})}\BibitemShut {NoStop}%
\bibitem [{\citenamefont {Bulla}\ \emph {et~al.}(2008)\citenamefont {Bulla},
  \citenamefont {Costi},\ and\ \citenamefont {Pruschke}}]{bulla2008}%
  \BibitemOpen
  \bibfield  {author} {\bibinfo {author} {\bibfnamefont {R.}~\bibnamefont
  {Bulla}}, \bibinfo {author} {\bibfnamefont {T.~A.}\ \bibnamefont {Costi}},\
  and\ \bibinfo {author} {\bibfnamefont {T.}~\bibnamefont {Pruschke}},\
  }\bibfield  {title} {\bibinfo {title} {Numerical renormalization group method
  for quantum impurity systems},\ }\href
  {https://doi.org/10.1103/RevModPhys.80.395} {\bibfield  {journal} {\bibinfo
  {journal} {Rev. Mod. Phys.}\ }\textbf {\bibinfo {volume} {80}},\ \bibinfo
  {pages} {395} (\bibinfo {year} {2008})}\BibitemShut {NoStop}%
\bibitem [{\citenamefont {Weichselbaum}\ and\ \citenamefont {von
  Delft}(2007)}]{weichselbaum2007}%
  \BibitemOpen
  \bibfield  {author} {\bibinfo {author} {\bibfnamefont {A.}~\bibnamefont
  {Weichselbaum}}\ and\ \bibinfo {author} {\bibfnamefont {J.}~\bibnamefont {von
  Delft}},\ }\bibfield  {title} {\bibinfo {title} {Sum-rule conserving spectral
  functions from the numerical renormalization group},\ }\href
  {https://doi.org/10.1103/PhysRevLett.99.076402} {\bibfield  {journal}
  {\bibinfo  {journal} {Phys. Rev. Lett.}\ }\textbf {\bibinfo {volume} {99}},\
  \bibinfo {pages} {076402} (\bibinfo {year} {2007})}\BibitemShut {NoStop}%
\bibitem [{\citenamefont {Horodecki}\ \emph {et~al.}(1999)\citenamefont
  {Horodecki}, \citenamefont {Horodecki},\ and\ \citenamefont
  {Horodecki}}]{horodecki1999}%
  \BibitemOpen
  \bibfield  {author} {\bibinfo {author} {\bibfnamefont {M.}~\bibnamefont
  {Horodecki}}, \bibinfo {author} {\bibfnamefont {P.}~\bibnamefont
  {Horodecki}},\ and\ \bibinfo {author} {\bibfnamefont {R.}~\bibnamefont
  {Horodecki}},\ }\bibfield  {title} {\bibinfo {title} {General teleportation
  channel, singlet fraction, and quasidistillation},\ }\href
  {https://doi.org/10.1103/PhysRevA.60.1888} {\bibfield  {journal} {\bibinfo
  {journal} {Phys. Rev. A}\ }\textbf {\bibinfo {volume} {60}},\ \bibinfo
  {pages} {1888} (\bibinfo {year} {1999})}\BibitemShut {NoStop}%
\bibitem [{\citenamefont {Wootters}(1998)}]{wootters1998}%
  \BibitemOpen
  \bibfield  {author} {\bibinfo {author} {\bibfnamefont {W.~K.}\ \bibnamefont
  {Wootters}},\ }\bibfield  {title} {\bibinfo {title} {Entanglement of
  formation of an arbitrary state of two qubits},\ }\href
  {https://doi.org/10.1103/PhysRevLett.80.2245} {\bibfield  {journal} {\bibinfo
   {journal} {Phys. Rev. Lett.}\ }\textbf {\bibinfo {volume} {80}},\ \bibinfo
  {pages} {2245} (\bibinfo {year} {1998})}\BibitemShut {NoStop}%
\bibitem [{\citenamefont {Verstraete}\ and\ \citenamefont
  {Verschelde}(2002)}]{verstraete2002}%
  \BibitemOpen
  \bibfield  {author} {\bibinfo {author} {\bibfnamefont {F.}~\bibnamefont
  {Verstraete}}\ and\ \bibinfo {author} {\bibfnamefont {H.}~\bibnamefont
  {Verschelde}},\ }\bibfield  {title} {\bibinfo {title} {Fidelity of mixed
  states of two qubits},\ }\href {https://doi.org/10.1103/PhysRevA.66.022307}
  {\bibfield  {journal} {\bibinfo  {journal} {Phys. Rev. A}\ }\textbf {\bibinfo
  {volume} {66}},\ \bibinfo {pages} {022307} (\bibinfo {year}
  {2002})}\BibitemShut {NoStop}%
\bibitem [{\citenamefont {Jones}\ \emph {et~al.}(1988)\citenamefont {Jones},
  \citenamefont {Varma},\ and\ \citenamefont {Wilkins}}]{jones1988}%
  \BibitemOpen
  \bibfield  {author} {\bibinfo {author} {\bibfnamefont {B.~A.}\ \bibnamefont
  {Jones}}, \bibinfo {author} {\bibfnamefont {C.~M.}\ \bibnamefont {Varma}},\
  and\ \bibinfo {author} {\bibfnamefont {J.~W.}\ \bibnamefont {Wilkins}},\
  }\bibfield  {title} {\bibinfo {title} {Low-temperature properties of the
  two-impurity {K}ondo {H}amiltonian},\ }\href
  {https://doi.org/10.1103/PhysRevLett.61.125} {\bibfield  {journal} {\bibinfo
  {journal} {Phys. Rev. Lett.}\ }\textbf {\bibinfo {volume} {61}},\ \bibinfo
  {pages} {125} (\bibinfo {year} {1988})}\BibitemShut {NoStop}%
\bibitem [{\citenamefont {Werner}(1989)}]{werner1989}%
  \BibitemOpen
  \bibfield  {author} {\bibinfo {author} {\bibfnamefont {R.~F.}\ \bibnamefont
  {Werner}},\ }\bibfield  {title} {\bibinfo {title} {Quantum states with
  {E}instein-{P}odolsky-{R}osen correlations admitting a hidden-variable
  model},\ }\href {https://doi.org/10.1103/PhysRevA.40.4277} {\bibfield
  {journal} {\bibinfo  {journal} {Phys. Rev. A}\ }\textbf {\bibinfo {volume}
  {40}},\ \bibinfo {pages} {4277} (\bibinfo {year} {1989})}\BibitemShut
  {NoStop}%
\bibitem [{\citenamefont {Charnes}\ and\ \citenamefont
  {Cooper}(1962)}]{charnes1962}%
  \BibitemOpen
  \bibfield  {author} {\bibinfo {author} {\bibfnamefont {A.}~\bibnamefont
  {Charnes}}\ and\ \bibinfo {author} {\bibfnamefont {W.~W.}\ \bibnamefont
  {Cooper}},\ }\bibfield  {title} {\bibinfo {title} {Programming with linear
  fractional functionals},\ }\href@noop {} {\bibfield  {journal} {\bibinfo
  {journal} {Naval Res. Logist. Q.}\ }\textbf {\bibinfo {volume} {9}},\
  \bibinfo {pages} {181} (\bibinfo {year} {1962})}\BibitemShut {NoStop}%
\bibitem [{\citenamefont {Boyd}\ and\ \citenamefont
  {Vandenberghe}(2004)}]{boyd2004}%
  \BibitemOpen
  \bibfield  {author} {\bibinfo {author} {\bibfnamefont {S.}~\bibnamefont
  {Boyd}}\ and\ \bibinfo {author} {\bibfnamefont {L.}~\bibnamefont
  {Vandenberghe}},\ }\href@noop {} {\emph {\bibinfo {title} {Convex
  Optimization}}}\ (\bibinfo  {publisher} {Cambridge University Press},\
  \bibinfo {address} {Cambridge},\ \bibinfo {year} {2004})\BibitemShut
  {NoStop}%
\bibitem [{\citenamefont {Backus}(1970)}]{backus1970}%
  \BibitemOpen
  \bibfield  {author} {\bibinfo {author} {\bibfnamefont {G.}~\bibnamefont
  {Backus}},\ }\bibfield  {title} {\bibinfo {title} {Inference from inadequate
  and inaccurate data, {I}},\ }\href@noop {} {\bibfield  {journal} {\bibinfo
  {journal} {Proc. Natl. Acad. Sci. USA}\ }\textbf {\bibinfo {volume} {65}},\
  \bibinfo {pages} {1} (\bibinfo {year} {1970})}\BibitemShut {NoStop}%
\bibitem [{\citenamefont {Stark}(1992)}]{stark1992}%
  \BibitemOpen
  \bibfield  {author} {\bibinfo {author} {\bibfnamefont {P.~B.}\ \bibnamefont
  {Stark}},\ }\bibfield  {title} {\bibinfo {title} {Inference in
  infinite-dimensional inverse problems: discretization and duality},\
  }\href@noop {} {\bibfield  {journal} {\bibinfo  {journal} {J. Geophys. Res.}\
  }\textbf {\bibinfo {volume} {97}},\ \bibinfo {pages} {14055} (\bibinfo {year}
  {1992})}\BibitemShut {NoStop}%
\bibitem [{\citenamefont {\v{S}id\'ak}(1967)}]{sidak1967}%
  \BibitemOpen
  \bibfield  {author} {\bibinfo {author} {\bibfnamefont {Z.}~\bibnamefont
  {\v{S}id\'ak}},\ }\bibfield  {title} {\bibinfo {title} {Rectangular
  confidence regions for the means of multivariate normal distributions},\
  }\href@noop {} {\bibfield  {journal} {\bibinfo  {journal} {J. Am. Stat.
  Assoc.}\ }\textbf {\bibinfo {volume} {62}},\ \bibinfo {pages} {626} (\bibinfo
  {year} {1967})}\BibitemShut {NoStop}%
\bibitem [{\citenamefont {\v{S}afr\'anek}(2017)}]{safranek2017}%
  \BibitemOpen
  \bibfield  {author} {\bibinfo {author} {\bibfnamefont {D.}~\bibnamefont
  {\v{S}afr\'anek}},\ }\bibfield  {title} {\bibinfo {title} {Discontinuities of
  the quantum {F}isher information and the {B}ures metric},\ }\href
  {https://doi.org/10.1103/PhysRevA.95.052320} {\bibfield  {journal} {\bibinfo
  {journal} {Phys. Rev. A}\ }\textbf {\bibinfo {volume} {95}},\ \bibinfo
  {pages} {052320} (\bibinfo {year} {2017})}\BibitemShut {NoStop}%
\bibitem [{\citenamefont {T\'oth}\ and\ \citenamefont
  {Apellaniz}(2014)}]{toth2014}%
  \BibitemOpen
  \bibfield  {author} {\bibinfo {author} {\bibfnamefont {G.}~\bibnamefont
  {T\'oth}}\ and\ \bibinfo {author} {\bibfnamefont {I.}~\bibnamefont
  {Apellaniz}},\ }\bibfield  {title} {\bibinfo {title} {Quantum metrology from
  a quantum information science perspective},\ }\href
  {https://doi.org/10.1088/1751-8113/47/42/424006} {\bibfield  {journal}
  {\bibinfo  {journal} {J. Phys. A}\ }\textbf {\bibinfo {volume} {47}},\
  \bibinfo {pages} {424006} (\bibinfo {year} {2014})}\BibitemShut {NoStop}%
\bibitem [{\citenamefont {\v{Z}itko}\ and\ \citenamefont
  {Pruschke}(2009)}]{zitko2009}%
  \BibitemOpen
  \bibfield  {author} {\bibinfo {author} {\bibfnamefont {R.}~\bibnamefont
  {\v{Z}itko}}\ and\ \bibinfo {author} {\bibfnamefont {T.}~\bibnamefont
  {Pruschke}},\ }\bibfield  {title} {\bibinfo {title} {Energy resolution and
  discretization artifacts in the numerical renormalization group},\ }\href
  {https://doi.org/10.1103/PhysRevB.79.085106} {\bibfield  {journal} {\bibinfo
  {journal} {Phys. Rev. B}\ }\textbf {\bibinfo {volume} {79}},\ \bibinfo
  {pages} {085106} (\bibinfo {year} {2009})}\BibitemShut {NoStop}%
\end{thebibliography}%

\clearpage
\onecolumngrid
\makeatletter\let\addcontentsline\qfi@addcontentsline\makeatother
\begin{center}{\large\bfseries Supplemental Material for ``Certifying Nonlocal Spin Quantum Entanglement from Local Tunneling Spectra''}\end{center}
\vspace{1ex}
\setcounter{equation}{0}\setcounter{figure}{0}\setcounter{table}{0}\setcounter{section}{0}
\setcounter{secnumdepth}{2}
\renewcommand{\thesection}{S\arabic{section}}
\renewcommand{\theequation}{S\arabic{equation}}
\renewcommand{\thetable}{S\arabic{table}}
\renewcommand{\thefigure}{S\arabic{figure}}
\makeatletter\def\theHsection{S\arabic{section}}\def\theHequation{S\arabic{equation}}\def\theHtable{S\arabic{table}}\def\theHfigure{S\arabic{figure}}\makeatother

\tableofcontents

\section{System, notation and assumptions}
\label{sec:S1}
We consider $N$ sites with spin operators $\bm S_i$ acting on $\mathcal H_i$ ($\dim\mathcal H_i=2S+1$), coupled to an environment $E$ with Hilbert space $\mathcal H_E$, which may contain substrate electrons, nuclear spins or unresolved magnetic neighbors. All quantities refer to the full state $\rho$ on $\mathcal H=\bigotimes_i\mathcal H_i\otimes\mathcal H_E$.

For $\rho=\sum_m\lambda_m|m\rangle\langle m|$ and Hermitian operators $G_1,\dots,G_M$ the quantum Fisher information (QFI) matrix is~\cite{braunstein1994,safranek2017}
\begin{equation}
F_{ij}[\rho]=2\sum_{m,n:\,\lambda_m+\lambda_n>0}\frac{(\lambda_m-\lambda_n)^2}{\lambda_m+\lambda_n}\,\mathrm{Re}\big[(G_i)_{mn}(G_j)_{nm}\big],
\label{eq:qfi}
\end{equation}
and $F_Q[\rho,G]$ denotes the diagonal element for a single generator $G$. We use the following standard properties~\cite{braunstein1994,pezze2018,toth2014}. The QFI of a linear combination is the quadratic form $F_Q[\rho,\sum_iw_iG_i]=\bm w^TF\bm w$, and the matrix $F$ is positive semidefinite. The QFI is convex in $\rho$. For pure states $F_Q=4\,\mathrm{Var}(G)$, while for mixed states $F_Q\le4\,\mathrm{Var}(G)$. For product states and generators acting on separate factors, the QFI is additive.

We denote by $K^a_{ij}$ the QFI matrix of the operators $\{S_i^a\}_{i=1}^N$ for spin axis $a\in\{x,y,z\}$ and set $K=\sum_aK^a$. Its diagonal elements $D_i=K_{ii}$ are the local QFIs, $\epsilon=\bm1^TK\bm1=\sum_aF_Q[\rho,S_{\rm tot}^a]$ is the leakage of the total spin, and $W(\bm w)=\bm w^TK\bm w$ is the collective QFI. For a dimer $W\equiv W(1,-1)$. The same quantity evaluated for the reduced two-spin state $\rho_{12}=\Tr_E\rho$ is denoted by $W_{\rm cl}$.

The results rely on the following physical assumptions.
\begin{itemize}
\item[(i)] \emph{Cluster identification.} The probed sites are the $N$ spin-$S$ moments of the cluster, located on distinct atoms or molecular units, and no other spin exchange-coupled to them is omitted. Section~\ref{sec:S6} shows that this assumption cannot be replaced by any property of the local spectra.
\item[(ii)] \emph{Fixed local moments.} $\bm S_i^2=S(S+1)$ in all populated states, i.e., charge fluctuations are negligible (Sec.~\ref{sec:S8} quantifies deviations).
\item[(iii)] \emph{Weak spin-exchange tunneling.} Tunneling is treated to second order, with amplitudes $\tau_0,\tau_1$ that do not depend on energy within the bias window and flat electrode densities of states. The tunneling current does not drive the spins out of equilibrium. Electrodes are unpolarized unless stated otherwise (Sec.~\ref{sec:S3}).
\item[(iv)] \emph{Thermal equilibrium.} $\rho=e^{-\beta H}/Z$ for the full Hamiltonian, with a spin temperature $T_s=1/\beta$ within a known interval $[T_s^-,T_s^+]$.
\item[(v)] \emph{Instrumental broadening.} Lock-in modulation of amplitude $V_a$, Gaussian voltage noise of width $\sigma_V$, a bias offset $V_0$ and the electron temperature $T_e$ are known or lie within stated ranges.
\item[(vi)] \emph{Measurement noise.} The errors at the $n_b$ bias points are independent and Gaussian with known standard deviations $\sigma_b$ (Sec.~\ref{sec:S7} discusses relaxations).
\end{itemize}

\section{Entanglement criteria for spins coupled to an environment}
\label{sec:S2}
\subsection{Separable states of spins and environment}
Suppose first that spins and environment are in a fully separable state, $\rho=\sum_kp_k\,\sigma_1^k\otimes\cdots\otimes\sigma_N^k\otimes\sigma_E^k$. Since the QFI is convex, it suffices to consider pure product states $|\psi\rangle=\bigotimes_i|\psi_i\rangle\otimes|\phi_E\rangle$, for which $F_Q=4\,\mathrm{Var}$ and the variances of independent factors add,
\begin{equation}
W(\bm w)=4\sum_a\mathrm{Var}\Big(\sum_iw_iS_i^a\Big)=4\sum_iw_i^2\big[S(S+1)-|\langle\bm S_i\rangle|^2\big].
\label{eq:prodvar}
\end{equation}
For $S=\frac12$ every pure state is a spin coherent state with $|\langle\bm S\rangle|=\frac12$, whereas for $S\ge1$ the maximum $S(S+1)$ is reached by states with $\langle\bm S\rangle=0$, e.g.\ $(|S\rangle+|{-S}\rangle)/\sqrt2$. Hence
\begin{equation}
W(\bm w)\le B_3(S)|\bm w|^2,\qquad B_3(\tfrac12)=2,\quad B_3(S\ge1)=4S(S+1).
\label{eq:T1}
\end{equation}
For a spin-$\frac12$ dimer and $\bm w=(1,-1)$ this gives $W\le4$.

\subsection{Spins entangled with private environments}
Next we allow each spin to be entangled with its own part of the environment. Such states have the form $\rho=\sum_kp_k\bigotimes_j\sigma_j^k$, where $\sigma_j^k$ is a state of spin $j$ together with an environment part $E_j$, for some partition $\mathcal H_E=\bigotimes_j\mathcal H_{E_j}$. This situation includes Kondo screening of each spin by an independent electron bath. For pure states $\bigotimes_j|\phi_j\rangle$ with $|\phi_j\rangle\in\mathcal H_j\otimes\mathcal H_{E_j}$ the variances again add, and $\sum_a\mathrm{Var}_{\phi_j}(S_j^a)\le\langle\bm S_j^2\rangle=S(S+1)$, so that by convexity
\begin{equation}
W(\bm w)\le4S(S+1)|\bm w|^2 .
\label{eq:T2}
\end{equation}
For the spin-$\frac12$ dimer this is $W\le6$. The bound is independent of how the environment is partitioned, so its violation excludes a state of this form for every way of sharing the environment. For fermionic environments the partition refers to sets of electronic modes and the states respect fermion-parity superselection. Local observables of even parity then factorize, and the argument is unchanged. Since $\mathrm{Var}(S^z)\le S^2$ in any state, the single-axis analog
\begin{equation}
F_Q\Big[\rho,\sum_iw_iS_i^z\Big]\le4S^2|\bm w|^2
\label{eq:U1}
\end{equation}
holds for both types of separable states.

\subsection{Entanglement of the two spins themselves}
For every state of two spin-$\frac12$ sites and an arbitrary environment, the QFI is bounded by the variance, $F_Q[\rho,S_1^a-S_2^a]\le4\langle(S_1^a-S_2^a)^2\rangle$. Summing over the three axes and using $\bm S_1\cdot\bm S_2=\frac14-P_s$, with the singlet projector $P_s=|\psi^-\rangle\langle\psi^-|$, we obtain
\begin{equation}
W\le4\langle(\bm S_1-\bm S_2)^2\rangle=4\big[\tfrac32-2\langle\bm S_1\cdot\bm S_2\rangle\big]=4(1+2a),\qquad a=\langle\psi^-|\rho_{12}|\psi^-\rangle .
\label{eq:T3}
\end{equation}
Separable two-spin states have a singlet fraction $a\le\frac12$~\cite{horodecki1999}, so $W>8$ implies that $\rho_{12}$ is entangled. Moreover, the fully entangled fraction $F\ge a$ obeys $F\le(1+C)/2$, with $C$ the concurrence~\cite{wootters1998,verstraete2002}, which yields the quantitative statement
\begin{equation}
C(\rho_{12})\ge2a-1\ge\frac{W-8}{4}.
\label{eq:conc}
\end{equation}
For spin $S$, $4\langle(\bm S_1-\bm S_2)^2\rangle=4[2S(S+1)-2\langle\bm S_1\cdot\bm S_2\rangle]$, and separable states satisfy $\langle\bm S_1\cdot\bm S_2\rangle=\sum_kp_k\langle\bm S_1\rangle_k\cdot\langle\bm S_2\rangle_k\ge-S^2$. Hence $W>8S(2S+1)$ implies entanglement of $\rho_{12}$.

\subsection{States that reach the bounds}
The three thresholds $4<6<8$ of the spin-$\frac12$ dimer are reached by explicit states. The N\'eel product state $|{\uparrow\downarrow}\rangle$ has $D_i=2$, $\epsilon=4$ and $W=4$. Two spins that each form a singlet with a private partner, $|\psi^-\rangle_{S_1a_1}\otimes|\psi^-\rangle_{S_2a_2}$ with $a_j\in E_j$, give $W=6$, while $\rho_{12}=\mathbb1/4$. The bath-mediated state $(|{\uparrow\downarrow}\rangle|e_0\rangle+|{\downarrow\uparrow}\rangle|e_1\rangle)/\sqrt2$ with orthonormal environment states gives $D_i=3$, $\epsilon=4$, $W=8$, $a=\frac12$ and a separable $\rho_{12}$. The criteria form a hierarchy. Fully separable states are a special case of private-environment states (take $E_1=E$). Tracing out the environment from a private-environment state leaves a separable $\rho_{12}$. Therefore $W>8$ implies $W>6$-type entanglement, which in turn implies violation of Eq.~\eqref{eq:T1}. The product bound~\eqref{eq:T1} addresses intersite entanglement only if the spins are known not to be entangled with the environment, an assumption violated by Kondo screening. The other two criteria require no such knowledge.

\section{Spin-exchange tunneling spectroscopy}
\label{sec:S3}
\subsection{Tunneling rates}
We describe tunneling between tip and sample by $H_T=\sum_{\sigma\sigma'}c^\dagger_{t\sigma}[\tau_0\delta_{\sigma\sigma'}+\tau_1\bm S_i\cdot\bm\sigma_{\sigma\sigma'}]c_{s\sigma'}+\mathrm{h.c.}$ and characterize each electrode by a spin density matrix $\mathcal D=\mathbb1+P\,\hat{\bm n}\cdot\bm\sigma$ for its occupied and empty states. To second order, with flat densities of states, the differential conductance is
\begin{equation}
\frac{dI}{dV}\propto\sum_{mn}p_n\Big\{\tfrac12\Tr[\mathcal D_s^{\rm emp}X_{mn}\mathcal D_t^{\rm occ}X_{mn}^\dagger]\,\Phi'_T(eV-\Delta_{mn})+\tfrac12\Tr[\mathcal D_t^{\rm emp}X_{mn}\mathcal D_s^{\rm occ}X_{mn}^\dagger]\,\Phi'_T(-eV-\Delta_{mn})\Big\},
\label{eq:rates}
\end{equation}
where $X_{mn}=\tau_0\delta_{mn}\mathbb1+\tau_1\langle m|\bm S_i|n\rangle\cdot\bm\sigma$, $\Delta_{mn}=E_m-E_n$, and $\Phi_T(x)=x/(1-e^{-x/T})$ results from the energy integral over the Fermi functions. Its second derivative $\Phi_T''$ is the autocorrelation of the derivative of the Fermi function and hence non-negative.

For unpolarized electrodes, $\frac12\Tr[X_{mn}X_{mn}^\dagger]=\tau_0^2\delta_{mn}+\tau_1^2\sum_a|\langle m|S_i^a|n\rangle|^2$, since the cross terms contain $\Tr\,\sigma^a=0$. Adding the two transfer directions yields the thermal step $\theta_T$, and the identity $\Phi'_T(x)+\Phi'_T(-x)=1$ gives the constant $\tau_0^2$ term. The result is Eq.~(1) of the main text.

\subsection{Spin-polarized electrodes}
If only the tip is polarized, Eq.~\eqref{eq:rates} acquires the additional terms $2P\tau_0\tau_1\delta_{mn}\hat{\bm n}\cdot\mathrm{Re}\langle n|\bm S_i|n\rangle+iP\tau_1^2\,\hat{\bm n}\cdot(\bm s_{mn}^*\times\bm s_{mn})$, with $\bm s_{mn}=\langle m|\bm S_i|n\rangle$. After summation over initial states with weights $p_n$ and over all transitions with the same energy, both terms are components of vectors constructed from $\rho$ and $H$ alone. For a state and Hamiltonian invariant under spin rotations such vectors vanish, so a polarized tip on an unpolarized substrate leaves the conductance unchanged. In a magnetic field this cancellation fails, as expected.

If both electrodes are polarized, the energy-resolved sum $\sum p_n s^a_{mn}s^{b*}_{mn}$ over a shell of fixed transition energy reduces to $\delta_{ab}S_i(E)/3$ for rotation-invariant states. With the identity $\sum_a\sigma^aA\sigma^a=2\Tr(A)\mathbb1-A$ for $2\times2$ matrices one finds $\sum_a\Tr[\sigma^a\mathcal D_s\sigma^a\mathcal D_t]=2(3-\Pi)$ and $\Tr[\mathcal D_s\mathcal D_t]=2(1+\Pi)$, with $\Pi=P_sP_t\,\hat{\bm m}\cdot\hat{\bm n}$ for the relevant pair of occupied and empty states. Hence, for each transfer direction $d$,
\begin{equation}
\frac{dI}{dV}\propto\sum_{d}\Big[\tau_0^2(1+\Pi_d)\Phi'_T(s_deV)+\tau_1^2\big(1-\tfrac{\Pi_d}{3}\big)\!\int\! S_i(E)\Phi_T'(s_deV-E)\,dE\Big],\qquad s_d=\pm1 .
\end{equation}
If the polarizations are the same for both directions, this is Eq.~(1) with $\tau_0^2\to\tau_0^2(1+\Pi)$ and $\tau_1^2\to\tau_1^2(1-\Pi/3)$. All bounds remain valid with $r^2\to r^2(1+\Pi)/(1-\Pi/3)$.

\subsection{Properties of the broadened thermal kernel}
Let $\theta(\omega,E)=\Phi'_{T_e}(\omega-E)+\Phi'_{T_e}(-\omega-E)$ and let $\langle\cdot\rangle$ denote the average over the lock-in modulation, the voltage noise and the bias offset. We use the following properties. (a) For $E\ge0$, $\mathrm{th}_1(E)\equiv\langle\theta(\omega,E)\rangle$ decreases and $\mathrm{th}_2(E)\equiv\langle\theta(\omega,-E)\rangle$ increases with $E$, since $\partial_E\theta(\omega,\pm E)=\mp[\Phi''(\omega\mp E)+\Phi''(-\omega\mp E)]$ and $\Phi''\ge0$. Moreover, $\mathrm{th}_2\ge\mathrm{th}_1$ and $\mathrm{th}_{1,2}(0)=1$. (b) With $g(z)=z/(1-e^{-z})$ one has $0\le\Phi''_T\le1/(6T)$ and $|\Phi'''_T|\le C_3/T^2$, where $C_3=0.045$ exceeds the maximum $0.04237$ of $|g'''|$ at $|z|=2.144$. (c) Consequently the composite kernel $\Theta=\mathrm{th}_1+e^{-\beta E}\mathrm{th}_2$ has a bounded slope, $|\partial_E\Theta|\le1/(3T_e)+e^{-\beta E}[2\beta+1/(3T_e)]$. (d) The Fourier transform of $\Phi''_T$ is $[\pi kT/\sinh(\pi kT)]^2>0$, that of the Gaussian noise is positive, and that of the semicircular lock-in kernel, $2J_1(kV_a)/(kV_a)$, has only isolated zeros.

\section{Local QFI from the tunneling spectrum}
\label{sec:S4}
\subsection{Fluctuation-dissipation relation for the QFI}
For a thermal state, $p_m=p_ne^{-\beta\Delta_{mn}}$, and the weights in Eq.~\eqref{eq:qfi} become $(p_n-p_m)^2/(p_n+p_m)=p_n\tanh(\beta\Delta_{mn}/2)(1-e^{-\beta\Delta_{mn}})$. For any Hermitian $O$ with spectral function $S_O(\omega)=\sum_{mn}p_n|O_{mn}|^2\delta(\omega-\Delta_{mn})$ this gives~\cite{hauke2016,shitara2016}
\begin{equation}
F_Q[\rho,O]=2\int d\omega\,\tanh\tfrac{\beta\omega}{2}\big(1-e^{-\beta\omega}\big)S_O(\omega)=4\int_0^\infty d\omega\,w_\beta(\omega)S_O(\omega),
\label{eq:thermalS}
\end{equation}
where the second form follows from detailed balance, $S_O(-\omega)=e^{-\beta\omega}S_O(\omega)$. Summing over the three components of $\bm S_i$ gives Eq.~(2) of the main text. The second form is also the numerically robust one. Weights at negative frequencies enter the first form multiplied by $e^{\beta|\omega|}$, so any truncation or inaccuracy of these small weights is strongly amplified. Under assumption (ii) the spectral weight obeys the sum rule $\int S_i(E)\,dE=\sum_a\langle(S_i^a)^2\rangle=S(S+1)$, and we shall repeatedly use the identity $w_\beta(E)=\tanh^2(\beta E/2)(1+e^{-\beta E})$.

\subsection{Scale-free ratio}
Writing the conductance in units of $\kappa\tau_1^2$ as $g(V)=b+\int\Theta(eV;E)\,dX(E)$, with $X=S_i|_{E>0}$ and $b=r^2+W_{\rm el}$, we use $\theta(\omega,0)=1$ for the elastic weight and detailed balance to fold $E<0$ onto $E>0$. The denominator of the ratio $R$ [Eq.~(3) of the main text] is then $r^2+W_{\rm el}+\int_{E>0}[S_i(E)+S_i(-E)]\,dE=r^2+S(S+1)$, and by Eq.~\eqref{eq:thermalS} its numerator is $S(S+1)D_i$, which proves Eq.~(4) of the main text,
\begin{equation}
R=\frac{D_i}{1+r^2/S(S+1)} .
\label{eq:prop1S}
\end{equation}
Since $R$ is unchanged when $b$ and $X$ are multiplied by the same positive factor, the minimum of $R$ over all pairs $(b,X)$ that reproduce the data up to an unknown overall scale is a lower bound on $D_i$.

\subsection{Finite bias window}
In practice the data cover $|eV|\le V_{\max}$, and the spectral weight is represented up to an energy $E_{\rm cut}\ge V_{\max}+V_a+mT_e$. Weight above $E_{\rm cut}$ contributes inside the window a constant $2e^{-\beta E}$ per unit weight, up to corrections smaller than $4\Phi'_{T_e}(-mT_e)$ ($<10^{-7}$ for $m=20$) that we add to the tolerances. The pair $(b+\int_{E>E_{\rm cut}}2e^{-\beta E}dX,\ X|_{E\le E_{\rm cut}})$ therefore reproduces the data. With $A=S(S+1)$, the weight measure $\delta=\int_{E>E_{\rm cut}}(1-e^{-\beta E})\,dX$ and $D^{\rm win}=4\int_{E\le E_{\rm cut}}w_\beta\,dX$, its ratio is
\begin{equation}
R'=\frac{A\,D^{\rm win}}{r^2+A-\delta} .
\label{eq:window}
\end{equation}
Because $w_\beta\le1+e^{-\beta E}$ and $\int_{E\le E_{\rm cut}}(1+e^{-\beta E})\,dX\le A-\delta$, we have $D^{\rm win}\le4(A-\delta)$ and therefore $R'-D^{\rm win}\le D^{\rm win}\delta/(A-\delta)\le4\delta$. On the other hand, $D_i-D^{\rm win}=4\int_{E>E_{\rm cut}}w_\beta\,dX\ge4\tanh(\beta E_{\rm cut}/2)\,\delta$. Combining both inequalities,
\begin{equation}
R'\le D_i+4\Big[1-\tanh\frac{\beta E_{\rm cut}}{2}\Big]\delta\le D_i+8e^{-\beta E_{\rm cut}}\delta .
\label{eq:windowbound}
\end{equation}
The bound is reached for $r=0$ at low temperature when all in-window weight lies far above $T$ and the elastic weight vanishes. Unmeasured high-energy weight is thus compensated by the scale-free normalization at an exponentially small cost, and the spectrum need not lie entirely within the window (Table~\ref{tab:window}).

\section{From local to collective spin fluctuations}
\label{sec:S5}
Since $K$ is positive semidefinite with $K_{ii}=D_i$ and $\bm1^TK\bm1=\epsilon$, the collective QFI of a dimer satisfies $W(1,-1)+W(1,1)=2(D_1+D_2)$ with $W(1,1)=\epsilon$, which is Eq.~(5) of the main text.

\subsection{Optimal bound for a dimer}
We seek the smallest largest eigenvalue $\lambda_{\max}(K)$ compatible with $K_{11}\ge D_1\low$, $K_{22}\ge D_2\low$ and $\bm1^TK\bm1\le\epsilon_{\max}$. Relaxing the leakage can only lower the minimum, so we set $\epsilon=\epsilon_{\max}$ and write $K=\begin{pmatrix}a&c\\c&b\end{pmatrix}$. The largest eigenvalue $\frac12[a+b+\sqrt{(a-b)^2+4c^2}]$ grows with $|c|$. The constraint $a+b+2c\le\epsilon$ allows $c=0$ if $a+b\le\epsilon$. Otherwise the smallest admissible $|c|$ is $(a+b-\epsilon)/2$. This value is compatible with positivity, $c^2\le ab$, only if $(\sqrt a-\sqrt b)^2\le\epsilon$. Hence $\lambda_{\max}\ge\tilde\lambda(a,b)$ with $\tilde\lambda=\max(a,b)$ for $a+b\le\epsilon$ and $\tilde\lambda=\frac12[a+b+\sqrt{(a-b)^2+(a+b-\epsilon)^2}]$ otherwise. The function $\tilde\lambda$ is continuous and non-decreasing in $a$ and $b$: for $a+b>\epsilon$, $\partial_a\tilde\lambda=\frac12[1+(u+v)/\sqrt{u^2+v^2}]\ge0$ with $u=a-b$ and $v=a+b-\epsilon\ge0$, because $u+v<0$ requires $u<0\le v$, and then $(u+v)^2\le u^2+v^2$. All admissible pairs satisfy $a\ge a^*$ and $b\ge b^*$, with
$a^*=\max[D_1\low,(\sqrt{D_2\low}-\sqrt{\epsilon})_+^2]$ and $b^*=\max[D_2\low,(\sqrt{D_1\low}-\sqrt{\epsilon})_+^2]$, and $(a^*,b^*)$ is itself admissible. The optimal bound is therefore
\begin{equation}
\lambda^*=\begin{cases}\max(a^*,b^*), & a^*+b^*\le\epsilon_{\max},\\[2pt] \tfrac12\Big[a^*+b^*+\sqrt{(a^*-b^*)^2+(a^*+b^*-\epsilon_{\max})^2}\Big], & \text{otherwise.}\end{cases}
\label{eq:dimeropt}
\end{equation}
If upper bounds $D_i^{\rm up}$ are also available, the constraints are inconsistent when $a^*>D_1^{\rm up}$ or $b^*>D_2^{\rm up}$. Equation~\eqref{eq:dimeropt} agrees with a numerical semidefinite-programming solution to $1.5\times10^{-8}$.

\subsection{Single measured site}
Minimizing $W=2(D_1+D_2)-\epsilon$ over $D_2\ge0$ and $\epsilon\le\epsilon_{\max}$, subject to positivity, $(\sqrt{D_1}-\sqrt{D_2})^2\le\epsilon$, gives
\begin{equation}
W\ge\big(2\sqrt{D_1\low}-\sqrt{\epsilon_{\max}}\big)^2\quad(D_1\low\ge\epsilon_{\max}),
\label{eq:single}
\end{equation}
and $W\ge\max(0,2D_1\low-\epsilon_{\max})$ otherwise. The minimum is attained at $\sqrt{D_2}=\sqrt{D_1}-\sqrt\epsilon$ and $\epsilon=\epsilon_{\max}$.

\subsection{Larger clusters}
For $N$ sites and $\bm u=\bm1/\sqrt N$, $\Tr K-\bm u^TK\bm u=\Tr[(\mathbb1-\bm u\bm u^T)K]\le(N-1)\lambda_{\max}$, while $\Tr K=\sum_iD_i$ and $\bm u^TK\bm u=\epsilon/N$. Hence
\begin{equation}
\lambda_{\max}(K)\ge\frac{\sum_iD_i-\epsilon/N}{N-1} .
\label{eq:cluster}
\end{equation}
The optimal bound follows from semidefinite programming. One minimizes $t$ subject to $t\mathbb1-K\succeq0$, $K\succeq0$, $K_{ii}\ge D_i\low$ and $\bm1^TK\bm1\le\epsilon_{\max}$. Since $\max_{|\bm w|=1}W(\bm w)=\lambda_{\max}(K)$, a bound exceeding $B_3(S)$ or $4S(S+1)$ excludes the corresponding separable states of Sec.~\ref{sec:S2}. For three sites with $\epsilon=0$, $K$ is fixed by its diagonal, $K_{12}=(D_3-D_1-D_2)/2$ and cyclic permutations, and it is positive semidefinite if and only if the $\sqrt{D_i}$ satisfy the triangle inequalities, since $K$ is then the Gram matrix of three vectors summing to zero. This provides a consistency check of measured local QFIs.

If only $S^z_{\rm tot}$ is conserved (e.g., for XXZ exchange in a field), the single-axis relation $W^z(1,-1)=2(D_1^z+D_2^z)-\epsilon^z$ and the bound~\eqref{eq:U1} apply, the latter being automatically valid for private environments. Axis-resolved spectra require a probe polarized along $z$ and are not considered in the main text.

\section{Leakage of the total spin}
\label{sec:S6}
\subsection{General bound}
Since $p_n-p_m$ and $\Delta_{mn}$ have the same sign in a thermal state and $|\tanh x|\le|x|$, $(p_n-p_m)^2/(p_n+p_m)=(p_n-p_m)\tanh(\beta\Delta_{mn}/2)\le(p_n-p_m)\beta\Delta_{mn}/2$. Inserting this into Eq.~\eqref{eq:qfi} gives
\begin{equation}
F_Q[\rho_\beta,G]\le\beta\sum_{mn}(p_n-p_m)\Delta_{mn}|G_{mn}|^2=\beta\langle[G,[H,G]]\rangle_\beta .
\label{eq:dc}
\end{equation}
For $G=S^a_{\rm tot}$ the double commutator vanishes for isotropic intra-cluster exchange and is proportional to the anisotropic couplings otherwise.

\subsection{Kondo coupling to second order}
For $H=J\bm S_1\cdot\bm S_2+H_{\rm b}+J_K\sum_i\bm S_i\cdot\bm s_i(0)$ with independent baths and $T\ll J$, the singlet ground state acquires to first order the triplet admixture $-J_K\sum|t\rangle|{\rm ph}\rangle\langle t,{\rm ph}|\bm S_i\cdot\bm s_i|\psi^-,{\rm FS}\rangle/(J+E_{\rm ph})$, where $|{\rm ph}\rangle$ denotes particle-hole excitations of the Fermi sea. With the bath spin structure factor per component, $S_b(\omega)=\sum_{\rm ph}|\langle{\rm ph}|s_i^a(0)|{\rm FS}\rangle|^2\delta(\omega-E_{\rm ph})$, and $\sum_t\langle t|S_i^a|\psi^-\rangle\langle\psi^-|S_i^b|t\rangle=\delta_{ab}/4$, the triplet weight from both baths is $w_t=\frac32J_K^2\int S_b(\omega)(J+\omega)^{-2}d\omega$. For the pure ground state $\langle\bm S_{\rm tot}\rangle=0$ and $\epsilon=4\langle\bm S_{\rm tot}^2\rangle=8w_t$, so that
\begin{equation}
\epsilon_0=12J_K^2\int_0^\infty\frac{S_b(\omega)}{(J+\omega)^2}\,d\omega .
\label{eq:eps0}
\end{equation}
The triplet decays to the singlet by emitting a particle-hole pair of energy $J$ into either bath, with the golden-rule rate
\begin{equation}
\Gamma_E=2\pi J_K^2\cdot2\cdot\tfrac14\,S_b(J)=\pi J_K^2S_b(J).
\label{eq:gamma}
\end{equation}
For a flat band with density of states $\nu$ per spin and particle-hole range $W_b$, $S_b(\omega)=\nu^2\omega/2$ and $\int_0^{W_b}\omega(J+\omega)^{-2}d\omega=\Lambda(W_b/J)$, which yields $\epsilon_0=6(J_K\nu)^2\Lambda=(12/\pi)(\Gamma_E/J)\Lambda$, Eq.~(6) of the main text. For a dimer coupled to discretized baths, the ratio of the exact leakage to Eq.~\eqref{eq:eps0} extrapolates to 0.999 for $J_K\to0$. With logarithmically discretized baths of two and four levels and bandwidths $J$ and $4J$, $\epsilon/\epsilon_{\rm PT}=1.01$--$1.03$ at $J_K=0.05J$ and $1.02$--$1.06$ at $J_K=0.1J$, approaching unity linearly (Table~\ref{tab:l2ed}).

For a flat band of half-width $W_b$ the particle-hole spectrum is $S_b(\omega)=\nu^2\omega/2$ only up to $\omega=W_b$ and then decreases linearly to zero at $2W_b$, which adds $W_b/(W_b+J)-\ln[(2W_b+J)/(W_b+J)]$ to $\Lambda$ and raises the second-order leakage above Eq.~(6) by 5--8\% for $W_b/J=100$--1000. At finite temperature, transitions within the thermally populated triplet raise it further, to 1.22--1.36 times Eq.~(6) at $T=0.5J$. Beyond second order, the NRG results of Sec.~\ref{sec:S8} exceed the exact second-order values $\epsilon_2$ by a factor that changes by less than 7\% between $T=0.02J$ and $0.5J$ and follows $[1-\nu J_K\ln(W_b/J)]^{-1}$ within 3\% for $\nu J_K\le0.05$ (Table~\ref{tab:l2nrg}), as expected from the logarithmic growth of the antiferromagnetic Kondo coupling toward low energies. Altogether $\epsilon=\xi\epsilon_0$, with $\epsilon_0$ evaluated at the golden-rule width, with $\xi=1.21$--1.24 for $\nu J_K=0.02$ and 1.44--1.64 for $\nu J_K=0.05$ at $T\le0.1J$, and 1.52--2.02 at $T=0.5J$. The same renormalization increases the triplet decay rate, so the measured $\Gamma_E$ is expected to exceed its golden-rule value at the bare coupling, in which case using the measured width in Eq.~(6) together with this $\xi$ is conservative. At these couplings the NRG lines are narrower than the numerical broadening. A direct test, which compares the spectral weight in windows next to the line between couplings, was inconclusive. For some discretization offsets the discretized line splits into peaks about $0.066J$ apart, and the raw spectra contain negative weights of up to $10^{-3}$, larger than the tails at weak coupling. The expectation therefore rests on a leading-logarithmic argument. The width is set by the coupling at the scale $J$, which grows as $[1-\nu J_K\ln(W_b/J)]^{-1}$ toward low energies, so that $\Gamma_E\approx\Gamma_E^{\rm GR}[1-\nu J_K\ln(W_b/J)]^{-2}\ge\Gamma_E^{\rm GR}$. The leakage grows only as $[1-\nu J_K\ln(W_b/J)]^{-1}$ (Sec.~\ref{sec:S8}), because it collects contributions from all scales between $J$ and $W_b$. Using the measured width together with $\xi$ is thus conservative at leading-logarithmic order. This is an expectation, not a numerical result.

For two spins that couple to one shared band, the local orbitals $\psi_{1,2}=a_ef_e(0)\pm a_of_o(0)$, with $a_{e,o}=\sqrt{(1\pm S)/2}$, overlap by $S$. Relative to independent bands, the coupling then contains $\bm S_{\rm tot}$ with weight $1+S^2$ and $\bm S_1-\bm S_2$ with weight $1-S^2$. The golden-rule width scales as $1-S^2$, and we define $\xi=\epsilon/[(1-S^2)\epsilon_0]$. At second order $\xi$ does not depend on $S$ for $T\ll J$ ($\xi_2=1.068$ at $W_b=250J$ and $T=0.02J$). At finite temperature, transitions within the triplet raise it to $\xi_2=1.23$, 1.32 and 2.07 for $S=0.3$, 0.6 and 0.9 at $T=0.3J$. Beyond second order the NRG results of Sec.~\ref{sec:S8} add a factor that grows with $S$. At $\nu J_K=0.05$, $W_b=250J$ and $T=0.02J$, $\xi/\xi_2=1.41$, 1.46, 1.60 and 2.10 for $S=0$, 0.3, 0.6 and 0.9 (Table~\ref{tab:l2nrg_shared}). For a shared band the measured width therefore bounds the leakage only together with the $\xi$ of the largest plausible overlap.

\subsection{Zero temperature}
If $\rho=|\Psi\rangle\langle\Psi|$ with $|\Psi\rangle$ a total spin singlet of spins and environment, the Wigner-Eckart theorem gives $\langle\bm S_i\rangle=0$, and the pure-state QFI equals four times the variance. Hence $D_i=4\langle\bm S_i^2\rangle=3$, $W=4\langle(\bm S_1-\bm S_2)^2\rangle=4(\frac32-2\langle\bm S_1\cdot\bm S_2\rangle)$ and $\epsilon=4(\frac32+2\langle\bm S_1\cdot\bm S_2\rangle)$. Hence
\begin{equation}
D_i=3,\qquad W=12-\epsilon,\qquad a=\tfrac14-\langle\bm S_1\cdot\bm S_2\rangle=1-\frac\epsilon8 .
\label{eq:t0S}
\end{equation}
Rotation invariance of $|\Psi\rangle$ makes $\rho_{12}$ invariant under $U\otimes U$, i.e., a Werner state $a|\psi^-\rangle\langle\psi^-|+(1-a)P_t/3$, which is entangled if and only if $a>\frac12$~\cite{werner1989}. From Eq.~\eqref{eq:qfi},
\begin{equation}
W_{\rm cl}=g(\epsilon)\equiv\frac{4(3-\epsilon/2)^2}{3-\epsilon/4}=12-3\epsilon+\frac{\epsilon^2}{12}+O(\epsilon^3).
\label{eq:gS}
\end{equation}
Thus $\rho_{12}$ is entangled if and only if $\epsilon<4$, i.e., if and only if $W>8$, whereas $W>6$ holds for $\epsilon<6$. In the discretized-bath calculations of Sec.~\ref{sec:S8} these relations hold to $10^{-8}$ at $T=0.005J$ for 63 of 72 parameter sets. The exceptions occur at strong coupling and narrow bandwidth ($J_K\ge1.5J$, $W_b=J$), where the lowest excitation gap, $0.0006$--$0.08J$, becomes comparable to this temperature. Evaluated in the ground state, which remains a non-degenerate singlet, the relations hold to $10^{-13}$ (Table~\ref{tab:t0}). With continuous bands (NRG, Sec.~\ref{sec:S8}), $W+\epsilon=12$ holds to $1.4\times10^{-4}$ and $a=1-\epsilon/8$ to $1.2\times10^{-8}$ at $T=0.02J$ for $\nu J_K\le0.05$. With a shared band, $W+\epsilon=12$ holds to $2\times10^{-4}$ for $S\le0.6$ and to $1.7\times10^{-3}$ for $S=0.9$.

\subsection{Environment gap}
\label{sec:envgap}
The difference $\delta_{\rm env}=W-W_{\rm cl}\ge0$ is the part of $W$ carried by correlations between the spins and their environment. A bound $\delta_{\rm env}\le2\epsilon$ would imply $W_{\rm cl}\ge W-2\epsilon$ and thus entanglement of $\rho_{12}$ whenever $W-2\epsilon>4$ (Sec.~\ref{sec:S2}). We summarize what can be established about such a bound. The results support it at weak coupling but do not prove it at finite coupling, and the certificate does not use it.

At zero temperature, Eqs.~\eqref{eq:t0S} and \eqref{eq:gS} give exactly
\begin{equation}
\delta_{\rm env}=12-\epsilon-g(\epsilon)=2\epsilon-\frac{\epsilon^2}{12-\epsilon}
\label{eq:envgap0}
\end{equation}
for any non-degenerate singlet ground state, whatever the form and strength of the coupling. The discretized-bath results of Sec.~\ref{sec:S8} satisfy Eq.~\eqref{eq:envgap0} to $3\times10^{-11}$ in the 63 parameter sets that meet the zero-temperature criterion.

At finite temperature the bound does not hold in general. A coupling to the exchange energy, $\lambda\,\bm S_1\cdot\bm S_2\otimes O$ with $O$ an operator of the environment, as produced for instance by lattice vibrations that modulate $J$, commutes with $\bm S_{\rm tot}$, so that $\epsilon=0$, but it correlates the environment with the singlet and triplet sectors and thereby gives $\delta_{\rm env}>0$. For a two-level environment and $\lambda=0.05J$ we find $\delta_{\rm env}=1.5\times10^{-3}$ at $T=0.3J$. In the limit of a perfect record, the state $a|\psi^-\rangle\langle\psi^-|\otimes|0\rangle\langle0|+(1-a)\frac{P_t}{3}\otimes|1\rangle\langle1|$ with orthogonal environment states has $\epsilon=0$ and $W=4(1+2a)$, whereas its reduced Werner state has $W_{\rm cl}=4(4a-1)^2/(2a+1)$ and is separable for $a\le\frac12$. At $a=\frac12$, $W=8$ but $W_{\rm cl}=2$. Random searches over rotation-invariant thermal states of spins coupled to small environments also produced violations at strong coupling.

For couplings linear in the spins, $V=\lambda\sum_i\bm S_i\cdot\bm B_i$ with a rotation-invariant environment and $\langle\bm B_i\rangle=0$, the bound holds to leading order. This class includes Kondo exchange with independent, shared or correlated baths and hyperfine coupling to nuclear spins. Expanding in $\lambda$, $\delta_{\rm env}=\lambda^2\delta_2+O(\lambda^3)$ and $\epsilon=\lambda^2\epsilon_2+O(\lambda^3)$ with
\begin{equation}
0\le\delta_2\le2\epsilon_2
\label{eq:envgap2}
\end{equation}
at every temperature and for every spectrum of the environment, with strict inequality for $T>0$. Terms of the QFI linear in the change of the state depend only on its partial trace over the environment and cancel in $\delta_{\rm env}$. Since the first-order change $\rho_1$ has ${\rm Tr}_E\rho_1=0$, the second order is $\langle X,\mathcal R_0^{-1}X\rangle$ with $X=i[G,\rho_1]-\frac12\{\rho_1,L_0\}$, where $L_0$ is the symmetric logarithmic derivative of the uncoupled state $\rho_0$ and $\mathcal R_0Y=\frac12\{\rho_0,Y\}$. Writing $V=\lambda(\bm S_{\rm tot}\cdot\bm B_++\bm G_-\cdot\bm B_-)$ with $\bm G_-=\bm S_1-\bm S_2$ and $\bm B_\pm=(\bm B_1\pm\bm B_2)/2$, the two terms contribute on disjoint blocks, and rotation invariance reduces each to a sum over single transitions of the environment, of energy $\Omega\ge0$, weighted by the spectral function of $\bm B_\pm$. For one transition, with $x=e^{-\beta J}$, $y=e^{-\beta\Omega}$, $s_A=1+xy$, $s_B=x+y$ and the logarithmic means $a=L(1,xy)$, $b=L(x,y)$ and $c=L(1,y)$, where $L(p,q)=(p-q)/\ln(p/q)$, the $\bm B_-$ and $\bm B_+$ terms contribute to $2\epsilon_2-\delta_2$ positive multiples of
\begin{equation}
R_-=\frac{(as_B+bs_A)^2}{s_As_B(s_A+s_B)}-\frac{2x(a^2+b^2)}{(1+x)(s_A+s_B)},\qquad R_+=\frac{2xc^2[y(1-x)+x^2(1+y+xy+y^2)]}{(1+x)(1+y)s_As_B}.
\end{equation}
Because $\ln(\sinh t/t)-t$ decreases for $t>0$, $1\le a/b\le e^{\beta\min(J,\Omega)}\le1/x$, which gives $R_-\ge2ab/(s_A+s_B)>0$, while $R_+\ge0$ is manifest. These closed forms agree with exact diagonalization of small environments to seven digits. The ratio $\delta_2/2\epsilon_2$ approaches 1 for $T\to0$ at fixed $\Omega>0$ and about $\frac12$ for $T\gg J$, and it is small for elastic transitions ($\Omega\ll T\ll J$). The discretized-bath ratios $\delta_{\rm env}/2\epsilon$ at $J_K=0.05J$ follow this trend, decreasing from 0.99996--0.99999 at $T=0.005J$ to 0.25--0.62 at $T=0.5J$. With continuous bands (NRG) at $T=0.02J$, $\delta_{\rm env}$ follows Eq.~\eqref{eq:envgap0} to $1.4\times10^{-4}$ for $\nu J_K\le0.05$.

Equation~\eqref{eq:envgap2} is a weak claim. It does not control the remainder, because the third-order terms, which contain the Kondo logarithms, are nonzero. Nor does it control the limit $T\to0$, where the margin $2\epsilon_2-\delta_2$ vanishes and only Eq.~\eqref{eq:envgap0} applies. It also requires a coupling linear in the spins and exact rotation invariance, which excludes exchange modulation, magnetic fields and magnetic anisotropy. Together with the discretized-bath calculations, in which $\delta_{\rm env}\le2\epsilon$ holds in all 360 finite-temperature configurations up to $J_K=4J$ (Table~\ref{tab:l2ed}), for the two-level baths of Table~\ref{tab:leakage} at all 90 NRG points with independent continuous bands (Table~\ref{tab:l2nrg}) and at all reliable entries with a shared band (Table~\ref{tab:l2nrg_shared}), it suggests that $W-2\epsilon>4$ indicates entanglement of weakly coupled spin pairs. It is not a certificate, and the two-spin criterion of the Letter remains $W>8$.

\subsection{Privately screened spins}
If each spin forms a singlet with a private partner, every local spectrum coincides with that of the isolated dimer at the same $J$ and $T$, while $\epsilon=6R$, $W=6R$ and $\rho_{12}=\mathbb1/4$. If the partners are not recognized as part of the cluster, the leakage inferred from a dimer-like linewidth is too small and the naive estimate $W\low=12R$ exceeds every threshold. Because no property of the local spectra distinguishes the two situations, the cluster must be identified by independent means, such as spatial maps, atom manipulation, or spectra recorded on the partners.

\section{Lower bounds from noisy spectra}
\label{sec:S7}
\subsection{Discretized spectral weight and kernel bounds}
We divide $[0,E_{\rm cut}]$ into intervals $I_k=[E_k,E_{k+1})$ of width $h$ on $[0,2.4J]$ and $8h$ beyond, with $E_{\rm cut}=\max|eV|+V_a+20T_e$ (Sec.~\ref{sec:S4}). As variables we use $x_k=\int_{I_k}(1+q)\,dX$ with $q=e^{-\beta E}$, i.e., the weight of each interval together with its detailed-balance partner. The broadened data then read $\bar g(V_b)=b+\sum_k\int_{I_k}\tilde\Theta(V_b;E)(1+q)\,dX$ with the normalized kernel $\tilde\Theta=(\mathrm{th}_1+q\,\mathrm{th}_2)/(1+q)$, the denominator of $R$ is exactly $b+\sum_kx_k$, and the numerator weights $\omega_k=4S(S+1)\tanh^2(\beta_-E_k/2)$ with $\beta_-=1/T_s^+$ bound $4S(S+1)\tanh^2(\beta E/2)$ from below within each interval and for all admissible spin temperatures.

Within an interval $[e_-,e_+]$ and for $\beta\in[\beta_-,\beta_+]$, the monotonicity of Sec.~\ref{sec:S3}C implies, with $q_{\min}=e^{-\beta_+e_+}$ and $q_{\max}=e^{-\beta_-e_-}$,
\begin{equation}
\frac{\mathrm{th}_1(e_+)+q_{\min}\mathrm{th}_2(e_-)}{1+q_{\min}}\le\tilde\Theta(V;E)\le\frac{\mathrm{th}_1(e_-)+q_{\max}\mathrm{th}_2(e_+)}{1+q_{\max}} ,
\label{eq:mono}
\end{equation}
because $\tilde\Theta$ increases with $q$ when $\mathrm{th}_2\ge\mathrm{th}_1$. A second, independent enclosure follows from the curvature bound
\begin{equation}
|\partial_E^2\tilde\Theta|\le M_2(E)=\frac{2C_3}{T_e^2}+e^{-\beta E}\Big[\frac{4C_3}{T_e^2}+\frac{4\beta}{3T_e}+2\beta^2\Big],
\label{eq:curv}
\end{equation}
obtained by writing $\tilde\Theta=\mathrm{th}_1+\Delta\,s$ with $\Delta=\mathrm{th}_2-\mathrm{th}_1$ and $s=1/(1+e^{\beta E})$. Linear interpolation between the ends of an interval of length $\ell$ then deviates from $\tilde\Theta$ by at most $M_2\ell^2/8$. We use the tighter of the two enclosures on four sub-intervals per interval and obtain bounds $A_{bk}^-\le\tilde\Theta(V_b;E)\le A_{bk}^+$ for all $E\in I_k$. The lock-in average is evaluated with 12-node Gauss-Chebyshev quadrature, whose error decreases exponentially with the number of nodes for these analytic integrands. The difference to a 6-node rule, below $10^{-15}$ for the parameters used, is added as a margin.

\subsection{Optimization}
The lower bound is the minimum of $\sum_k\omega_kx_k/(b+\sum_kx_k)$ over $b,x_k\ge0$ subject to $b+\sum_kA_{bk}^-x_k\le y_b+\delta_b$ and $b+\sum_kA_{bk}^+x_k\ge y_b-\delta_b$ at every bias point. Introducing $t=1/(b+\sum_kx_k)$ and the rescaled variables $u_0=tb$, $u_k=tx_k$~\cite{charnes1962} turns this ratio into the linear program
\begin{equation}
D_i\low=\min\sum_k\omega_ku_k\quad\text{with}\quad u_0+\sum_kA^-_{bk}u_k\le t(y_b+\delta_b),\ \ u_0+\sum_kA^+_{bk}u_k\ge t(y_b-\delta_b),\ \ u_0+\sum_ku_k=1,\ \ u_0,u_k,t\ge0,
\label{eq:lp}
\end{equation}
which we solve numerically~\cite{boyd2004}. An independent lower estimate from the dual problem agrees with the optimum to $3\times10^{-10}$ in all cases studied. For any spectrum compatible with Eq.~(1) of the main text, the discretized true spectrum satisfies the constraints whenever all errors lie within the tolerances, and its discretized ratio does not exceed $R$, because the $\omega_k$ underestimate the numerator while the denominator is exact.

\subsection{Confidence level}
For independent Gaussian errors we set $\delta_b=z\sigma_b+\eta_w$, where $\eta_w$ accounts for the finite-window correction of Sec.~\ref{sec:S4} and $z$ solves $[2\Phi_{\mathcal N}(z)-1]^{n_b}=1-\alpha'$~\cite{sidak1967}, $\Phi_{\mathcal N}$ being the standard normal distribution function. With probability $1-\alpha'$ all errors lie within the tolerances, and then $D_i\low\le R'\le D_i+8e^{-\beta E_{\rm cut}}\delta$. The last term vanishes if the spectrum lies below $E_{\rm cut}$ and is below $10^{-15}$ otherwise. For independent data on $N$ sites and $\alpha'=\alpha/N$, all $N$ bounds hold simultaneously with probability at least $(1-\alpha/N)^N\ge1-\alpha$, and $W\low=2(D_1\low+D_2\low)-\epsilon_{\max}\le W$ whenever $\epsilon\le\epsilon_{\max}$.

\subsection{Sharpness}
When the bias points become dense and the tolerances and $h$ vanish, the bound converges to Eq.~\eqref{eq:prop1S} (or to $R'$ for spectra extending beyond the window). The true spectrum is always admissible, which gives the upper limit. Conversely, optimal solutions form a sequence of non-negative measures on a compact energy interval, and any limit reproduces $\bar g$ on a dense set of bias points and hence on the whole window. Since $\bar g$ is analytic in $V$ (the kernels are analytic in the strip $|\mathrm{Im}\,\omega|<2\pi T_e$), it is determined on the whole real axis. Its derivative is the convolution of the lock-in kernel, the noise distribution and $\Phi''$ with the odd measure $\tanh(\beta E/2)[\delta_E-\delta_{-E}]$ weighted by the spectral weights, because $\partial_\omega[\theta(\omega,E)+q\,\theta(\omega,-E)]=(1-q)[\Phi''(\omega-E)-\Phi''(\omega+E)]$. The Fourier transforms of the first three factors vanish at most at isolated points (Sec.~\ref{sec:S3}C), so the spectral weights on $(0,E_{\rm cut}]$ are uniquely determined and the limit is the true spectrum.

\subsection{Instrumental parameters and numerical settings}
An interval of spin temperatures is treated exactly through Eqs.~\eqref{eq:mono} and \eqref{eq:curv}. The electron temperature, lock-in amplitude, bias offset and voltage noise enter either by enclosing the kernels over a grid of parameter values, with a margin between grid points, or by taking the smallest bound over the grid, which is valid if the true values lie on the grid. Unknown $\sigma_b$ estimated from repeated sweeps can be treated with Student-$t$ quantiles. Correlated noise requires whitening. Unless stated otherwise we use 181 equidistant bias points on $|eV|\le2.4J+0.6J+12T$, $V_a=0.02J$, energy resolution $h=\max(T/32,0.0025J)$, and noise $\sigma_b=\eta G_\infty$ with $\eta=0.5\%$ of the high-bias conductance $G_\infty$. Synthetic data are generated with a finer lock-in quadrature (96 nodes) than the one used in the analysis, to avoid artificially perfect agreement.

\section{Models and numerical results}
\label{sec:S8}
All analytical relations used in this work were checked numerically against independent calculations based on explicit tunneling rates, full density matrices, exact diagonalization, direct minimization and random sampling of quantum states.

\subsection{Isolated antiferromagnetic dimer}
For $H=J\bm S_1\cdot\bm S_2$ at temperature $T$, with $q=e^{J/T}$, the singlet and triplet populations are $p_s=q/(q+3)$ and $p_t=1/(q+3)$, and the local structure factor is $S_1(E)=\frac34p_s\delta(E-J)+\frac34p_t\delta(E+J)+\frac32p_t\delta(E)$. This gives $D_i=3R$, $\epsilon=0$ and $W=12R$ with $R=(q-1)^2/[(q+3)(q+1)]$. The relative step height is $f=\frac34(p_s-p_t)/(r^2+\frac34)$, and $\rho_{12}$ is a Werner state with $a=p_s$. The thresholds follow from $W=12R>4\Leftrightarrow2q^2-10q>0\Leftrightarrow q>5$, $W>6\Leftrightarrow q^2-8q-1>0\Leftrightarrow q>4+\sqrt{17}$ and $W>8\Leftrightarrow q^2-14q-3>0\Leftrightarrow q>7+\sqrt{52}$, that is, $T<0.621J$, $0.477J$ and $0.377J$. The reduced state $\rho_{12}$ is entangled for $p_s>\frac12\Leftrightarrow q>3$, that is, $T<0.910J$. For exact data the bound equals $12R/(1+4r^2/3)$, so a threshold $W_t$ is exceeded for $r^2<\frac34(12R/W_t-1)$, which at $T\to0$ gives $r^2<\frac32$, $\frac34$ and $\frac38$. Since the relative step height at $T\to0$ is $f=1/(1+4r^2/3)$, the bound then equals $12f$: the thresholds 4, 6 and 8 correspond to $f>1/3$, $1/2$ and $2/3$, a condition that can be read directly from the measured spectrum. With $\eta=0.5\%$ and $T\le0.3J$, certification of $W>8$ in at least $99\%$ of the realizations requires $r\le0.3$, i.e., $f\gtrsim0.9$ (Table~\ref{tab:l1main}). Tables~\ref{tab:tight} and \ref{tab:coverage} summarize the accuracy of the bound for exact spectra and its statistical behavior for noisy spectra.

\begin{table}[h!]
\caption{Accuracy of the lower bound for exact spectra of the isolated dimer (181 bias points, $V_a=0.02J$). $h$ is the energy resolution and $K$ the number of energy intervals. The deficit relative to Eq.~(4) of the main text decreases linearly with $h$ and grows with temperature as the thermal kernel flattens.}\label{tab:tight}
\begin{ruledtabular}
\begin{tabular}{ccccccc}
$T/J$ & $h$ & $K$ & $r$ & Eq.~(4) & bound & relative deficit \\ \hline
0.05 & $T/8$ & 429 & 0.0 & 3.00000 & 3.00000 & 9.6e-09 \\
0.05 & $T/8$ & 429 & 0.6 & 2.02703 & 2.02695 & 3.8e-05 \\
0.05 & $T/8$ & 429 & 1.2 & 1.02740 & 1.02732 & 7.6e-05 \\
0.05 & $T/32$ & 1714 & 0.0 & 3.00000 & 3.00000 & 2.6e-09 \\
0.05 & $T/32$ & 1714 & 0.6 & 2.02703 & 2.02701 & 8.3e-06 \\
0.05 & $T/32$ & 1714 & 1.2 & 1.02740 & 1.02738 & 1.7e-05 \\
0.05 & $T/64$ & 3428 & 0.0 & 3.00000 & 3.00000 & 1.3e-09 \\
0.05 & $T/64$ & 3428 & 0.6 & 2.02703 & 2.02702 & 4.0e-06 \\
0.05 & $T/64$ & 3428 & 1.2 & 1.02740 & 1.02739 & 8.2e-06 \\
0.2 & $T/8$ & 132 & 0.0 & 2.88165 & 2.86076 & 7.2e-03 \\
0.2 & $T/8$ & 132 & 0.6 & 1.94706 & 1.92357 & 1.2e-02 \\
0.2 & $T/8$ & 132 & 1.2 & 0.98687 & 0.96495 & 2.2e-02 \\
0.2 & $T/32$ & 525 & 0.0 & 2.88165 & 2.87631 & 1.9e-03 \\
0.2 & $T/32$ & 525 & 0.6 & 1.94706 & 1.94284 & 2.2e-03 \\
0.2 & $T/32$ & 525 & 1.2 & 0.98687 & 0.98379 & 3.1e-03 \\
0.2 & $T/64$ & 1049 & 0.0 & 2.88165 & 2.87896 & 9.3e-04 \\
0.2 & $T/64$ & 1049 & 0.6 & 1.94706 & 1.94509 & 1.0e-03 \\
0.2 & $T/64$ & 1049 & 1.2 & 0.98687 & 0.98563 & 1.3e-03 \\
0.5 & $T/8$ & 73 & 0.0 & 1.40509 & 1.26896 & 9.7e-02 \\
0.5 & $T/8$ & 73 & 0.6 & 0.94939 & 0.83536 & 1.2e-01 \\
0.5 & $T/8$ & 73 & 1.2 & 0.48120 & 0.39486 & 1.8e-01 \\
0.5 & $T/32$ & 287 & 0.0 & 1.40509 & 1.37004 & 2.5e-02 \\
0.5 & $T/32$ & 287 & 0.6 & 0.94939 & 0.92290 & 2.8e-02 \\
0.5 & $T/32$ & 287 & 1.2 & 0.48120 & 0.46371 & 3.6e-02 \\
0.5 & $T/64$ & 574 & 0.0 & 1.40509 & 1.38750 & 1.3e-02 \\
0.5 & $T/64$ & 574 & 0.6 & 0.94939 & 0.93660 & 1.3e-02 \\
0.5 & $T/64$ & 574 & 1.2 & 0.48120 & 0.47339 & 1.6e-02 \\
\end{tabular}
\end{ruledtabular}
\end{table}

\begin{table}[h!]
\caption{Statistical behavior for noisy spectra of the isolated dimer, with 100 noise realizations for each of the two sites, noise $\eta=0.5\%$ of $G_\infty$, $n_b=181$ bias points, joint confidence level 95\% ($\alpha'=0.025$ per site, $z=3.81$). Coverage is the fraction of realizations in which both local bounds hold. The last column gives the fractions in which $W\low$ exceeds 4, 6 and 8.}\label{tab:coverage}
\begin{ruledtabular}
\begin{tabular}{ccccccc}
$(T/J,\,r)$ & $D_i$ & Eq.~(4) & $\langle D_i\low\rangle\pm$ s.d. & range & coverage & $P(W\low>4,6,8)$ \\ \hline
(0.05, 0.5) & 3.0000 & 2.2500 & 2.2188$\pm$0.0078 & [2.1998, 2.2430] & 1.00 & 1.00, 1.00, 1.00 \\
(0.2, 0.8) & 2.8817 & 1.5548 & 1.4997$\pm$0.0109 & [1.4741, 1.5345] & 1.00 & 1.00, 0.48, 0.00 \\
(0.4, 0.3) & 1.8744 & 1.6736 & 1.5541$\pm$0.0164 & [1.5115, 1.6019] & 1.00 & 1.00, 1.00, 0.00 \\
\end{tabular}
\end{ruledtabular}
\end{table}

\subsection{Large-scale tests of coverage and power}
The guarantee of Sec.~\ref{sec:S7} was tested on a computing cluster over 945 settings of temperature ($T/J=0.03$--$0.5$), tunneling ratio ($r=0$--$1.2$), noise level ($\eta=0.25$--$1\%$), lock-in amplitude ($V_a=0.01$--$0.05J$) and number of bias points ($n_b=121$--$241$), with 2000 noise realizations per setting and site ($1.89\times10^6$ joint realizations). The joint coverage never fell below $0.9990\pm0.0007$, and its mean was $1.0000$. At most $1.1\%$ of the realizations admitted no solution, and no bound was issued for them. The power depends mainly on $T$ and $r$ and hardly on $V_a$ and $n_b$: in the reference configuration $W\low$ exceeds 8, 6 and 4 in at least $99\%$ of the realizations for $r\le0.3$, $0.6$ and $0.9$, respectively, at $T\le0.3J$, and the mean bound lies $1$--$8\%$ below Eq.~(4) of the main text (Table~\ref{tab:l1main}). To test robustness, the data were generated with misspecified instrumental parameters (Table~\ref{tab:l1nuis}). Even the analysis with nominal parameters kept a coverage of at least $0.996$. Enclosing the kernels over the parameter ranges kept at least $0.999$, and taking the smallest bound over a parameter grid kept $1.000$. Both had larger power than the nominal analysis, whose realizations more often admit no solution.

\begin{table}[h!]
\footnotesize
\caption{Coverage and power for the isolated dimer from 2000 noise realizations per setting{,} two sites each{,} in the reference configuration (noise $\eta=0.5\%${,} $V_a=0.02J${,} $n_b=181${,} joint confidence level 95\%). ``No solution'' is the fraction of realizations whose data admit no admissible spectrum{,} for which no bound is issued. Over all 945 settings ($T/J=0.03$--$0.5${,} $r=0$--$1.2${,} $\eta=0.25$--$1\%${,} $V_a=0.01$--$0.05J${,} $n_b=121$--$241${,} 1{,}890{,}000 realizations in total){,} the smallest coverage was $0.9990\pm0.0007$.}\label{tab:l1main}
\begin{ruledtabular}
\begin{tabular}{ccccccccc}
$T/J$ & $r$ & coverage & $P(W\low>4)$ & $P(W\low>6)$ & $P(W\low>8)$ & $\langle D_i\low\rangle/D_i$ & Eq.~(4) & no solution \\ \hline
0.03 & 0.0 & 1.0000 & 0.992 & 0.992 & 0.992 & 0.992 & 1.000 & 0.0085 \\
0.03 & 0.3 & 1.0000 & 1.000 & 1.000 & 1.000 & 0.884 & 0.893 & 0.0000 \\
0.03 & 0.6 & 1.0000 & 0.999 & 0.999 & 0.235 & 0.665 & 0.676 & 0.0010 \\
0.03 & 0.9 & 1.0000 & 1.000 & 0.000 & 0.000 & 0.469 & 0.481 & 0.0000 \\
0.03 & 1.2 & 1.0000 & 0.056 & 0.000 & 0.000 & 0.330 & 0.342 & 0.0005 \\
0.05 & 0.0 & 0.9995 & 0.996 & 0.996 & 0.996 & 0.991 & 1.000 & 0.0045 \\
0.05 & 0.3 & 1.0000 & 1.000 & 1.000 & 1.000 & 0.884 & 0.893 & 0.0000 \\
0.05 & 0.6 & 1.0000 & 1.000 & 1.000 & 0.185 & 0.665 & 0.676 & 0.0005 \\
0.05 & 0.9 & 1.0000 & 1.000 & 0.000 & 0.000 & 0.469 & 0.481 & 0.0000 \\
0.05 & 1.2 & 1.0000 & 0.038 & 0.000 & 0.000 & 0.330 & 0.342 & 0.0005 \\
0.1 & 0.0 & 1.0000 & 0.998 & 0.998 & 0.998 & 0.990 & 1.000 & 0.0025 \\
0.1 & 0.3 & 1.0000 & 1.000 & 1.000 & 1.000 & 0.882 & 0.893 & 0.0005 \\
0.1 & 0.6 & 1.0000 & 0.999 & 0.999 & 0.059 & 0.664 & 0.676 & 0.0010 \\
0.1 & 0.9 & 1.0000 & 1.000 & 0.000 & 0.000 & 0.468 & 0.481 & 0.0005 \\
0.1 & 1.2 & 1.0000 & 0.015 & 0.000 & 0.000 & 0.328 & 0.342 & 0.0010 \\
0.2 & 0.0 & 1.0000 & 1.000 & 1.000 & 1.000 & 0.982 & 1.000 & 0.0005 \\
0.2 & 0.3 & 1.0000 & 1.000 & 1.000 & 1.000 & 0.874 & 0.893 & 0.0000 \\
0.2 & 0.6 & 1.0000 & 0.999 & 0.999 & 0.000 & 0.657 & 0.676 & 0.0010 \\
0.2 & 0.9 & 1.0000 & 0.999 & 0.000 & 0.000 & 0.461 & 0.481 & 0.0010 \\
0.2 & 1.2 & 1.0000 & 0.000 & 0.000 & 0.000 & 0.323 & 0.342 & 0.0000 \\
0.3 & 0.0 & 1.0000 & 1.000 & 1.000 & 1.000 & 0.962 & 1.000 & 0.0005 \\
0.3 & 0.3 & 1.0000 & 1.000 & 1.000 & 1.000 & 0.856 & 0.893 & 0.0005 \\
0.3 & 0.6 & 1.0000 & 1.000 & 1.000 & 0.000 & 0.640 & 0.676 & 0.0000 \\
0.3 & 0.9 & 1.0000 & 1.000 & 0.000 & 0.000 & 0.446 & 0.481 & 0.0005 \\
0.3 & 1.2 & 1.0000 & 0.000 & 0.000 & 0.000 & 0.309 & 0.342 & 0.0005 \\
0.4 & 0.0 & 1.0000 & 0.999 & 0.999 & 0.000 & 0.934 & 1.000 & 0.0010 \\
0.4 & 0.3 & 1.0000 & 1.000 & 1.000 & 0.000 & 0.829 & 0.893 & 0.0000 \\
0.4 & 0.6 & 1.0000 & 1.000 & 0.000 & 0.000 & 0.616 & 0.676 & 0.0000 \\
0.4 & 0.9 & 1.0000 & 0.000 & 0.000 & 0.000 & 0.424 & 0.481 & 0.0000 \\
0.4 & 1.2 & 1.0000 & 0.000 & 0.000 & 0.000 & 0.289 & 0.342 & 0.0005 \\
0.5 & 0.0 & 1.0000 & 1.000 & 0.000 & 0.000 & 0.900 & 1.000 & 0.0000 \\
0.5 & 0.3 & 1.0000 & 1.000 & 0.000 & 0.000 & 0.797 & 0.893 & 0.0000 \\
0.5 & 0.6 & 1.0000 & 0.000 & 0.000 & 0.000 & 0.586 & 0.676 & 0.0000 \\
0.5 & 0.9 & 1.0000 & 0.000 & 0.000 & 0.000 & 0.397 & 0.481 & 0.0010 \\
0.5 & 1.2 & 1.0000 & 0.000 & 0.000 & 0.000 & 0.264 & 0.342 & 0.0000 \\
\end{tabular}
\end{ruledtabular}
\end{table}

\begin{table}[h!]
\footnotesize
\caption{Coverage and power under instrumental misspecification (isolated dimer, $r=0.5$, $\eta=0.5\%$, $n_b=181$, 1000 realizations per setting, two sites each). The data were generated with electron temperature $0.9$--$1.1$ times, and lock-in amplitude $0.9$ or $1.1$ times, the nominal values, bias offset $0$ or $0.004J$, Gaussian voltage noise $0$ or $0.004J$ and a linear background slope $0$ or $0.01$. The analysis used the nominal values, or enclosed the kernels over $T_e$ and $V_a\pm10\%$, $|V_0|\le0.005J$, $\sigma_V\le0.004J$, $T_s\pm10\%$ and a background slope of up to $0.02$, or took the smallest bound over a grid of these values.}\label{tab:l1nuis}
\begin{ruledtabular}
\begin{tabular}{lccccccc}
method & settings & min.\ coverage & mean coverage & $P(W\low>4)$ & $P(W\low>6)$ & $P(W\low>8)$ & min.\ coverage at $T/J=0.05,\,0.2,\,0.4$ \\ \hline
nominal parameters & 144 & 0.9960 & 0.9999 & 0.930 & 0.596 & 0.596 & 0.9960, 1.0000, 1.0000 \\
kernels enclosed over ranges & 144 & 0.9990 & 1.0000 & 0.889 & 0.667 & 0.463 & 0.9990, 1.0000, 0.9990 \\
smallest bound over grid & 144 & 1.0000 & 1.0000 & 0.975 & 0.667 & 0.555 & 1.0000, 1.0000, 1.0000 \\
\end{tabular}
\end{ruledtabular}
\end{table}

\subsection{Finite bias window}
Table~\ref{tab:window} tests Eqs.~\eqref{eq:window} and \eqref{eq:windowbound} for a spectrum with a line at $E_1=J$ inside the window and a line at $E_2=8J$ above $E_{\rm cut}$.

\begin{table}[h!]
\caption{Finite bias window. The spectrum has a line of weight 0.3 at $E_1=J$ inside the window and a line of weight $x_2$ at $E_2=8J$ above $E_{\rm cut}$, at $T=0.1J$ with exact data. The optimized bound reproduces $R'$ of Eq.~\eqref{eq:window} and never exceeds $D_i$.}\label{tab:window}
\begin{ruledtabular}
\begin{tabular}{ccccc}
$x_2$ & $r$ & $D_i$ & $R'$ & bound \\ \hline
0.0 & 0.0 & 1.1998 & 1.1998 & 1.1995 \\
0.0 & 0.5 & 1.1998 & 0.8999 & 0.8995 \\
0.0 & 1.0 & 1.1998 & 0.5142 & 0.5139 \\
0.1 & 0.0 & 1.5998 & 1.3844 & 1.3841 \\
0.1 & 0.5 & 1.5998 & 0.9999 & 0.9995 \\
0.1 & 1.0 & 1.5998 & 0.5454 & 0.5451 \\
0.25 & 0.0 & 2.1998 & 1.7998 & 1.7994 \\
0.25 & 0.5 & 2.1998 & 1.1998 & 1.1995 \\
0.25 & 1.0 & 2.1998 & 0.5999 & 0.5996 \\
0.4 & 0.0 & 2.7998 & 2.5711 & 2.5708 \\
0.4 & 0.5 & 2.7998 & 1.4998 & 1.4995 \\
0.4 & 1.0 & 2.7998 & 0.6666 & 0.6662 \\
\end{tabular}
\end{ruledtabular}
\end{table}

\subsection{States that mimic entanglement}
Table~\ref{tab:controls} lists the exact QFIs of the test states discussed in the main text. In the ferromagnetic dimer ($J<0$) the triplet mixture gives $D_i=1$ and $W=4$. A naive analysis reads the N\'eel product state as $W=8$, but its leakage restores $W=4$. A spin-1 ion with single-ion anisotropy has a single step, which would be read as $W=12$ if the ion were mistaken for a spin-$\frac12$ dimer. The table also lists privately screened spins (Sec.~\ref{sec:S6}) and the bath-mediated state (Sec.~\ref{sec:S2}).

\begin{table}[h!]
\caption{Exact QFIs of the full state and of the reduced two-spin state for the test states of the main text. The naive estimate $2(D_1+D_2)$ ignores leakage. For the spin-1 ion it is the value obtained if its single spectral step were attributed to a spin-$\frac12$ dimer. The last column states whether the spins are entangled beyond what private environments allow (known by construction).}\label{tab:controls}
\begin{ruledtabular}
\begin{tabular}{lcccccccc}
state & $D_1$ & $\epsilon$ & $W$ & $W_{\rm cl}$ & $a$ & $C(\rho_{12})$ & naive & entangled \\ \hline
isolated dimer, $T=0.3J$ & 2.433 & 0.000 & 9.733 & 9.733 & 0.903 & 0.807 & 9.733 & yes \\
Kondo-coupled dimer, $J_K=0.1J$, $T\to0$ & 3.000 & 0.004 & 11.996 & 11.988 & 1.000 & 0.999 & 12.000 & yes \\
bath-mediated state & 3.000 & 4.000 & 8.000 & 4.000 & 0.500 & 0.000 & 12.000 & yes \\
privately screened spins & 3.000 & 6.000 & 6.000 & 0.000 & 0.250 & 0.000 & 12.000 & no \\
N\'eel product state & 2.000 & 4.000 & 4.000 & 4.000 & 0.500 & 0.000 & 8.000 & no \\
ferromagnetic dimer & 1.000 & 0.000 & 4.000 & 4.000 & -0.000 & 0.000 & 4.000 & no \\
spin-1 ion analysed as a dimer & -- & -- & -- & -- & -- & -- & 12.000 & -- \\
\end{tabular}
\end{ruledtabular}
\end{table}

\subsection{Charge fluctuations in a Hubbard dimer}
For the half-filled Hubbard dimer with $U\gg t$ the singlet-triplet gap is $J=(\sqrt{U^2+16t^2}-U)/2$, and the singlet ground state has a double occupancy $d$ per site, so that $\langle\bm s_1^2\rangle=\frac34(1-2d)$. Matrix elements of the vector operator $\bm s_1$ between singlets vanish, and ionic configurations carry no spin on site 1, so the covalent weight $1-2d$ of the ground state enters all spin transitions. The in-window spectrum is $\frac34(1-2d)p_s\delta(E-J)+\frac34(1-2d)p_t\delta(E+J)+\frac32p_t\delta(E)$, whereas the triplet-to-ionic weight $\frac32p_td$ lies near $U$, above the window. With $p_s+3p_t=1$ the high-bias plateau is $\frac34[1-2d(1-2p_t)]$, and for exact data and $r=0$ the bound becomes
\begin{equation}
D_1\low=\frac{3w_\beta(J)\,p_s(1-2d)}{1-2d(1-2p_t)} ,
\label{eq:hubbard}
\end{equation}
in agreement with the numerical optimization to $8\times10^{-13}$. At low temperature it exceeds the true $D_1$ by a factor approaching $(1-2d)^{-1}$: for $U/t=16$ ($d=0.0075$), $D_1/D_1\low=0.985$ at $T=0.02J$, while the ratio exceeds unity at higher temperatures. A conservative bound is obtained by multiplying $D_i\low$ by $1-2d_{\max}$.

\subsection{Kondo-coupled dimer}
Each spin couples with $J_K$ to the local orbital of its own electron bath, represented by two levels at $\pm0.5J$ (or $\pm1.5J$) with equal weights, i.e., four spin-orbitals per bath. The Hilbert space of dimension $4\times2^8=1024$ is diagonalized exactly, and three-axis quantities follow from the $z$ components by spin-rotation symmetry, which we verified against the full calculation. Table~\ref{tab:leakage} lists the results.

\begin{table}[h!]
\caption{Dimer Kondo-coupled to two discretized electron baths (two levels at $\pm0.5J$ or $\pm1.5J$ per bath), exact diagonalization. At $T=0.005J$ the relations $W=12-\epsilon$ and $W_{\rm cl}=g(\epsilon)$ [Eqs.~\eqref{eq:t0S} and \eqref{eq:gS}] hold to $10^{-8}$, and $\epsilon_{\rm PT}$ is the second-order result, Eq.~\eqref{eq:eps0}.}\label{tab:leakage}
\begin{ruledtabular}
\begin{tabular}{ccccccccccc}
levels & $J_K/J$ & $T/J$ & $\epsilon$ & $\epsilon_{\rm PT}$ & $W$ & $12-\epsilon$ & $W_{\rm cl}$ & $g(\epsilon)$ & $W-W_{\rm cl}$ & $2\epsilon$ \\ \hline
$\pm0.5$ & 0.05 & 0.005 & 0.0010 & 0.0009 & 11.9990 & 11.9990 & 11.9971 & 11.9971 & 0.0019 & 0.0019 \\
$\pm0.5$ & 0.05 & 0.1 & 0.0010 & 0.0009 & 11.9956 & 11.9990 & 11.9937 & 11.9970 & 0.0019 & 0.0020 \\
$\pm0.5$ & 0.05 & 0.25 & 0.0021 & 0.0009 & 10.7605 & 11.9979 & 10.7593 & 11.9938 & 0.0012 & 0.0041 \\
$\pm0.5$ & 0.1 & 0.005 & 0.0039 & 0.0038 & 11.9961 & 11.9961 & 11.9882 & 11.9882 & 0.0079 & 0.0079 \\
$\pm0.5$ & 0.1 & 0.1 & 0.0041 & 0.0038 & 11.9921 & 11.9959 & 11.9843 & 11.9876 & 0.0078 & 0.0083 \\
$\pm0.5$ & 0.1 & 0.25 & 0.0086 & 0.0038 & 10.7466 & 11.9914 & 10.7416 & 11.9743 & 0.0049 & 0.0171 \\
$\pm0.5$ & 0.3 & 0.005 & 0.0394 & 0.0338 & 11.9606 & 11.9606 & 11.8819 & 11.8819 & 0.0787 & 0.0788 \\
$\pm0.5$ & 0.3 & 0.1 & 0.0417 & 0.0338 & 11.9492 & 11.9583 & 11.8716 & 11.8751 & 0.0776 & 0.0834 \\
$\pm0.5$ & 0.3 & 0.25 & 0.0878 & 0.0338 & 10.5791 & 11.9122 & 10.5270 & 11.7373 & 0.0522 & 0.1755 \\
$\pm0.5$ & 0.6 & 0.005 & 0.1864 & 0.1350 & 11.8136 & 11.8136 & 11.4439 & 11.4439 & 0.3698 & 0.3727 \\
$\pm0.5$ & 0.6 & 0.1 & 0.2004 & 0.1350 & 11.7605 & 11.7996 & 11.3950 & 11.4023 & 0.3656 & 0.4007 \\
$\pm0.5$ & 0.6 & 0.25 & 0.4191 & 0.1350 & 9.9085 & 11.5809 & 9.6218 & 10.7579 & 0.2867 & 0.8382 \\
$\pm0.5$ & 1.0 & 0.005 & 0.6600 & 0.3750 & 11.3400 & 11.3400 & 10.0584 & 10.0584 & 1.2816 & 1.3200 \\
$\pm0.5$ & 1.0 & 0.1 & 0.7625 & 0.3750 & 11.0232 & 11.2375 & 9.7165 & 9.7642 & 1.3067 & 1.5250 \\
$\pm0.5$ & 1.0 & 0.25 & 1.4061 & 0.3750 & 8.2852 & 10.5939 & 6.9737 & 7.9684 & 1.3115 & 2.8122 \\
$\pm0.5$ & 1.7 & 0.005 & 2.7256 & 1.0837 & 9.2744 & 9.2744 & 4.6241 & 4.6241 & 4.6502 & 5.4513 \\
$\pm0.5$ & 1.7 & 0.1 & 3.6949 & 1.0837 & 7.4156 & 8.3051 & 2.3162 & 2.5592 & 5.0993 & 7.3898 \\
$\pm0.5$ & 1.7 & 0.25 & 3.8956 & 1.0837 & 6.6278 & 8.1044 & 1.7625 & 2.1857 & 4.8653 & 7.7912 \\
$\pm0.5$ & 2.2 & 0.005 & 4.2637 & 1.8150 & 7.7363 & 7.7363 & 1.5588 & 1.5588 & 6.1775 & 8.5274 \\
$\pm0.5$ & 2.2 & 0.1 & 4.8476 & 1.8150 & 6.7284 & 7.1524 & 0.6403 & 0.7427 & 6.0881 & 9.6952 \\
$\pm0.5$ & 2.2 & 0.25 & 4.8259 & 1.8150 & 6.6267 & 7.1741 & 0.6323 & 0.7686 & 5.9944 & 9.6518 \\
$\pm0.5$ & 3.0 & 0.005 & 5.2071 & 3.3750 & 6.7708 & 6.7929 & 0.3663 & 0.3702 & 6.4045 & 10.4141 \\
$\pm0.5$ & 3.0 & 0.1 & 5.3388 & 3.3750 & 6.5517 & 6.6612 & 0.2383 & 0.2625 & 6.3134 & 10.6776 \\
$\pm0.5$ & 3.0 & 0.25 & 5.3355 & 3.3750 & 6.5451 & 6.6645 & 0.2385 & 0.2650 & 6.3066 & 10.6710 \\
$\pm1.5$ & 0.5 & 0.005 & 0.0266 & 0.0234 & 11.9734 & 11.9734 & 11.9203 & 11.9203 & 0.0531 & 0.0532 \\
$\pm1.5$ & 0.5 & 0.1 & 0.0266 & 0.0234 & 11.9699 & 11.9734 & 11.9168 & 11.9203 & 0.0531 & 0.0532 \\
$\pm1.5$ & 0.5 & 0.25 & 0.0314 & 0.0234 & 10.7065 & 11.9686 & 10.6554 & 11.9060 & 0.0511 & 0.0627 \\
$\pm1.5$ & 1.0 & 0.005 & 0.1207 & 0.0938 & 11.8793 & 11.8793 & 11.6392 & 11.6392 & 0.2402 & 0.2414 \\
$\pm1.5$ & 1.0 & 0.1 & 0.1207 & 0.0938 & 11.8748 & 11.8793 & 11.6347 & 11.6390 & 0.2401 & 0.2415 \\
$\pm1.5$ & 1.0 & 0.25 & 0.1461 & 0.0938 & 10.4904 & 11.8539 & 10.2573 & 11.5635 & 0.2332 & 0.2922 \\
$\pm1.5$ & 2.0 & 0.005 & 0.6150 & 0.3750 & 11.3850 & 11.3850 & 10.1882 & 10.1882 & 1.1968 & 1.2300 \\
$\pm1.5$ & 2.0 & 0.1 & 0.6158 & 0.3750 & 11.3694 & 11.3842 & 10.1730 & 10.1859 & 1.1964 & 1.2316 \\
$\pm1.5$ & 2.0 & 0.25 & 0.8012 & 0.3750 & 9.2696 & 11.1988 & 8.0582 & 9.6538 & 1.2114 & 1.6023 \\
$\pm1.5$ & 3.5 & 0.005 & 2.3748 & 1.1484 & 9.6252 & 9.6252 & 5.4615 & 5.4615 & 4.1637 & 4.7497 \\
$\pm1.5$ & 3.5 & 0.1 & 2.4356 & 1.1484 & 9.2346 & 9.5644 & 5.1053 & 5.3135 & 4.1294 & 4.8712 \\
$\pm1.5$ & 3.5 & 0.25 & 3.2721 & 1.1484 & 6.0212 & 8.7279 & 1.9039 & 3.4105 & 4.1173 & 6.5442 \\
\end{tabular}
\end{ruledtabular}
\end{table}

Larger baths were treated with a logarithmic discretization of flat bands of half-width $W_b=J$ or $4J$: $M=2$ or 4 levels per bath at the midpoints of the intervals $[W_b\Lambda^{-(n+1)},W_b\Lambda^{-n}]$ and $[0,W_b\Lambda^{1-M/2}]$, with weights proportional to the interval widths ($\Lambda=2,3$). The Hamiltonian conserves the particle number of each bath and $S^z_{\rm tot}$. We diagonalize it exactly in these sectors with half-filled baths, up to dimension $19\,600$ for $M=4$, for $J_K/J=0.05$--$4$ and $T/J=0.005$--$0.5$ (72 parameter sets, Tables~\ref{tab:l2ed} and \ref{tab:t0}).

Continuous bands were treated with the full-density-matrix numerical renormalization group (NRG)~\cite{wilson1975,bulla2008,weichselbaum2007} of NRG Ljubljana~\cite{zitko2009}, for two spins each coupled to its own flat band of half-width $W_b$ with density of states $\nu=1/2W_b$ per spin, $\nu J_K=0.02$--0.2, $W_b/J=100$, 250 and 1000, and $T/J=0.02$--0.5 (90 points). We used the discretization parameter $\Lambda=3$, 12000 kept states with an energy cutoff of 14 times the running scale, and an average over four discretization offsets $z$. The quantities $D_i$, $\epsilon$ and $W$ follow from the raw spectral weights of $S_1^z$, $S_2^z$ and their cross correlator with the kernel of the QFI, and $W_{\rm cl}$ follows from the singlet weight $a$. The parallelogram relation $W+\epsilon=2(D_1+D_2)$ holds to $4\times10^{-15}$, detailed balance to $10^{-12}$, and the sum rule $3\sum_\omega A(\omega)=\frac34$ for $S_1^z$ to $4\times10^{-4}$ at $T\le0.1J$. At higher temperature the elastic weight of the thermally populated triplet, which carries no QFI, is not contained in the spectra, and the sum falls to 0.69. Increasing the number of kept states from 6000 to 12000 changes $\epsilon$ by 3.5\% at $\nu J_K=0.02$ and 0.5\% at 0.05, changing $\Lambda$ to 2.5 changes it by at most 0.2\%, and $D_i$ and $W$ change by less than $10^{-4}$. Small negative values of $W-W_{\rm cl}$, down to $-0.05$ at $\nu J_K=0.02$ and $T\ge0.3J$ where the exact value is of order 0.01, indicate an accuracy of about 1\% of $W$ at high temperature. Table~\ref{tab:l2nrg} lists the results at $T=0.02J$. For $\nu J_K\le0.1$, and for $\nu J_K=0.15$ with $W_b\le250J$, $\epsilon\le2.64$ and $W>8$. For $\nu J_K=0.2$, and for 0.15 with $W_b=1000J$, the spins are Kondo-screened, with $\epsilon=5.0$--5.9, $W_{\rm cl}=0.003$--0.55 and $W=6.1$--6.9. There $W>6$ persists up to $T=0.1$--$0.3J$. The line width $\Gamma_E$ is resolved only at these strong couplings, where it exceeds $J$. 

Dimers on one shared band were computed in the even and odd channels of Sec.~\ref{sec:S6} for overlaps $S=0.3$, 0.6 and 0.9. The channel-diagonal couplings are those of the independent model. The channel-mixing exchange $J_Ka_ea_o(\bm S_1-\bm S_2)\cdot(\bm s_{eo}+\bm s_{oe})$ breaks the separate charge conservation of the channels, and we added it to NRG Ljubljana as a custom model. Its impurity Hamiltonian agrees with an independent diagonalization to $2\times10^{-16}$. At $S=0$ the model reproduces the independent-band results to $6\times10^{-9}$ in $\epsilon$. We used 6000 kept states. With 12000 states $\epsilon$ changes by 1.2\% at $S=0.6$ and 1.6\% at $S=0.9$ ($\nu J_K=0.05$, $W_b=250J$, $T=0.02J$). The spin sum rule, checked for every $z$, fails for $S\ge0.99$. There the odd channel decouples, and the truncation no longer represents its states correctly. The sum rule also fails at strong mixing, namely for $\nu J_K\ge0.1$ with $S\ge0.6$, for $\nu J_K=0.05$ with $S=0.9$ at $T\ge0.1J$ and with $S=0.6$ at $T=0.3J$, and for $W_b=1000J$. We keep only the entries in which it holds to 0.01 (Table~\ref{tab:l2nrg_shared}). The leakage factor $\xi$ grows with $S$ at every coupling, as discussed in Sec.~\ref{sec:S6}. For $T\le0.1J$ and $\nu J_K\le0.05$ it reaches 2.24 at $S=0.9$, and at $T=0.3J$ it reaches 3.2. The zero-temperature relations and $W-W_{\rm cl}\le2\epsilon$ hold wherever the NRG is reliable. At strong coupling the shared band changes the ground state. For $S=0.6$ and $\nu J_K=0.2$ at $T=0.02J$, $W+\epsilon=11.1$ instead of 12, so the ground state is no longer a nondegenerate singlet, presumably because the even channel underscreens the triplet. There $W=4.6$, and no threshold is passed.

At $S=1$ the odd channel is absent, and the model reduces to two spins coupled to one channel. A one-channel calculation with the two-impurity model of NRG Ljubljana and the two-channel calculation at $S=1$ agree. At $\nu J_K=0.05$, $W_b=250J$ and $T=0.02J$ they give $\epsilon=0.1225$ and 0.1214 and $W=5.76$ and 5.86. The singlet weight is $a=0.67$ and $W+\epsilon=5.9$, so the state is not a decoupled singlet. The cause is the ferromagnetic indirect exchange between two spins that couple to the same orbital. It is of order $J_K^2\nu$, which for flat bands is $x=2(W_b/J)(\nu J_K)^2$ in units of $J$, and $x=1.25$ at this point. A one-channel scan over $x$ from 0.002 to 5 at $T=0.02J$ (Table~\ref{tab:s1c}) gives a decoupled singlet, with $W+\epsilon=12$ to $3\times10^{-10}$, at all nine points with $x\le0.5$. Points with equal $x$ but different $W_b$ and $\nu J_K$ agree, so $x$ is the controlling scale. At $x=1.25$ the lowest NRG level is still the singlet, but a sector with nonzero total spin is populated at this temperature, with $\langle\bm S_{\rm tot}^2\rangle=0.65$. For $x=2$ and 5 the ground state is a triplet that the single channel underscreens, with $\langle\bm S_{\rm tot}^2\rangle=2$, $a=0$ and $W=W_{\rm cl}=4$. At $x=1.25$ the reduced state remains entangled ($a>\frac12$), but $W$ passes only the weakest threshold. For $x\ge2$ no threshold is passed. The spin sum rule misses 0.05 to 0.06 of its weight $\frac14$ for some $z$, and only at the three points without a decoupled singlet. There the static $\langle\bm S_{\rm tot}^2\rangle$ also exceeds the binned spectral weight of $S_{\rm tot}$, by 0.34 to 0.66. Both reflect elastic ($\omega=0$) weight of the sector with nonzero total spin, which the binned spectra do not contain. At full overlap the singlet, and with it the certificate, therefore survives for $x\le0.5$ and is lost for $x\gtrsim1$. Overlaps between 0.9 and 1 were not computed.

\begin{table}[h!]
\footnotesize
\caption{Kondo-coupled dimer with logarithmically discretized baths of $M$ levels (discretization parameter $\Lambda=2$, with $\Lambda=3$ agreeing within 3\%), exact diagonalization in particle-number and spin sectors with half-filled baths. $\epsilon$, $W$ and $W_{\rm cl}$ at $T=0.005J$. Here $\epsilon_{\rm PT}$ is the second-order result. The environment gap stays below $2\epsilon$ at every temperature $T=0.05$--$0.5J$.}\label{tab:l2ed}
\begin{ruledtabular}
\begin{tabular}{ccccccccc}
$M$ & $W_b/J$ & $J_K/J$ & $\epsilon$ & $\epsilon/\epsilon_{\rm PT}$ & $W$ & $W_{\rm cl}$ & $\max_{T>0}(W-W_{\rm cl})/2\epsilon$ & $W$ at $T=0.3J$ \\ \hline
2 & 1 & 0.05 & 0.0010 & 1.025 & 11.9990 & 11.9971 & 1.000 & 9.729 \\
2 & 1 & 0.1 & 0.0039 & 1.052 & 11.9961 & 11.9882 & 1.000 & 9.717 \\
2 & 1 & 0.2 & 0.0166 & 1.108 & 11.9834 & 11.9502 & 0.999 & 9.665 \\
2 & 1 & 0.4 & 0.0740 & 1.233 & 11.9260 & 11.7785 & 0.997 & 9.434 \\
2 & 1 & 0.7 & 0.2690 & 1.464 & 11.7310 & 11.1992 & 0.989 & 8.744 \\
2 & 1 & 1 & 0.6600 & 1.760 & 11.3400 & 10.0584 & 0.971 & 7.742 \\
2 & 1 & 1.5 & 1.9871 & 2.355 & 10.0129 & 6.4330 & 0.893 & 6.418 \\
2 & 1 & 2.5 & 4.7678 & 2.034 & 7.2319 & 0.8397 & 0.615 & 6.446 \\
2 & 1 & 4 & 5.5338 & 0.922 & 6.4136 & 0.1266 & 0.566 & 6.395 \\
2 & 4 & 0.05 & 0.0002 & 1.010 & 11.9998 & 11.9995 & 1.000 & 9.733 \\
2 & 4 & 0.1 & 0.0006 & 1.020 & 11.9994 & 11.9982 & 1.000 & 9.732 \\
2 & 4 & 0.2 & 0.0025 & 1.041 & 11.9975 & 11.9925 & 1.000 & 9.728 \\
2 & 4 & 0.4 & 0.0104 & 1.084 & 11.9896 & 11.9688 & 1.000 & 9.711 \\
2 & 4 & 0.7 & 0.0338 & 1.151 & 11.9662 & 11.8986 & 0.999 & 9.662 \\
2 & 4 & 1 & 0.0734 & 1.223 & 11.9266 & 11.7804 & 0.997 & 9.578 \\
2 & 4 & 1.5 & 0.1823 & 1.350 & 11.8177 & 11.4559 & 0.992 & 9.344 \\
2 & 4 & 2.5 & 0.6098 & 1.626 & 11.3902 & 10.2034 & 0.973 & 8.427 \\
2 & 4 & 4 & 1.8947 & 1.974 & 10.1053 & 6.6711 & 0.906 & 6.157 \\
4 & 1 & 0.05 & 0.0011 & 1.027 & 11.9989 & 11.9968 & 0.999 & 9.728 \\
4 & 1 & 0.1 & 0.0044 & 1.056 & 11.9956 & 11.9869 & 0.999 & 9.714 \\
4 & 1 & 0.2 & 0.0185 & 1.117 & 11.9815 & 11.9445 & 0.999 & 9.652 \\
4 & 1 & 0.4 & 0.0833 & 1.258 & 11.9167 & 11.7506 & 0.996 & 9.382 \\
4 & 1 & 0.7 & 0.3117 & 1.536 & 11.6883 & 11.0731 & 0.985 & 8.590 \\
4 & 1 & 1 & 0.8064 & 1.947 & 11.1936 & 9.6389 & 0.959 & 7.528 \\
4 & 1 & 1.5 & 2.7089 & 2.907 & 9.2911 & 4.6631 & 0.783 & 6.427 \\
4 & 1 & 2.5 & 5.1190 & 1.978 & 6.8270 & 0.4429 & 0.615 & 6.553 \\
4 & 1 & 4 & 5.5616 & 0.839 & 6.4251 & 0.1171 & 0.567 & 6.414 \\
4 & 4 & 0.05 & 0.0002 & 1.012 & 11.9998 & 11.9994 & 1.000 & 9.733 \\
4 & 4 & 0.1 & 0.0008 & 1.024 & 11.9992 & 11.9976 & 1.000 & 9.731 \\
4 & 4 & 0.2 & 0.0033 & 1.049 & 11.9967 & 11.9900 & 1.000 & 9.725 \\
4 & 4 & 0.4 & 0.0140 & 1.102 & 11.9860 & 11.9581 & 0.999 & 9.700 \\
4 & 4 & 0.7 & 0.0461 & 1.186 & 11.9539 & 11.8619 & 0.998 & 9.623 \\
4 & 4 & 1 & 0.1014 & 1.278 & 11.8986 & 11.6967 & 0.996 & 9.490 \\
4 & 4 & 1.5 & 0.2586 & 1.449 & 11.7414 & 11.2300 & 0.989 & 9.109 \\
4 & 4 & 2.5 & 0.9142 & 1.844 & 11.0858 & 9.3327 & 0.959 & 7.639 \\
4 & 4 & 4 & 2.9169 & 2.298 & 9.0831 & 4.1861 & 0.837 & 5.549 \\
\end{tabular}
\end{ruledtabular}
\end{table}

\begin{table}[h!]
\footnotesize
\caption{Zero-temperature relations at strong coupling and narrow bandwidth. In all cases the ground state is a non-degenerate spin singlet and $W=12-\epsilon$ holds in the ground state to numerical precision. The deviations at $T=0.005J$ arise where the lowest excitation gap is comparable to that temperature.}\label{tab:t0}
\begin{ruledtabular}
\begin{tabular}{ccccccc}
$M$ & $W_b/J$ & $\Lambda$ & $J_K/J$ & gap$/J$ & $|W+\epsilon-12|$ at $T=0.005J$ & $|W+\epsilon-12|$ in the ground state \\ \hline
2 & 1 & 2 & 1 & 0.5684 & 5.3e-15 & 9e-15 \\
2 & 1 & 2 & 1.5 & 0.2946 & 1.2e-14 & 1e-14 \\
2 & 1 & 2 & 2.5 & 0.0463 & 2.8e-04 & 7e-15 \\
2 & 1 & 2 & 4 & 0.0060 & 5.3e-02 & 9e-15 \\
4 & 1 & 2 & 2.5 & 0.0111 & 5.4e-02 & 2e-14 \\
4 & 1 & 2 & 4 & 0.0013 & 1.3e-02 & 5e-15 \\
4 & 1 & 3 & 1.5 & 0.0815 & 3.4e-07 & 2e-14 \\
4 & 1 & 3 & 2.5 & 0.0051 & 4.4e-02 & 4e-14 \\
4 & 1 & 3 & 4 & 0.0006 & 5.8e-03 & 3e-14 \\
\end{tabular}
\end{ruledtabular}
\end{table}

\begin{table}[h!]
\footnotesize
\caption{Kondo-coupled dimer with two independent flat bands of half-width $W_b$ (density of states $\nu=1/2W_b$ per spin), full-density-matrix NRG (NRG Ljubljana, $\Lambda=3$, 12000 kept states, four values of $z$). All quantities at $T=0.02J$ except the last column. $\epsilon_0$ is Eq.~(6) of the main text with the golden-rule width $\Gamma_E=\frac{\pi}{2}(\nu J_K)^2J$, and $\epsilon_2$ the exact second-order leakage of these bands at the same temperature. $W-W_{\rm cl}\le2\epsilon$ at all 90 points ($T=0.02$--$0.5J$).}\label{tab:l2nrg}
\begin{ruledtabular}
\begin{tabular}{cccccccccc}
$\nu J_K$ & $W_b/J$ & $\epsilon$ & $W$ & $W_{\rm cl}$ & $\frac{W-W_{\rm cl}}{2\epsilon}$ & $\epsilon/\epsilon_0$ & $\epsilon/\epsilon_2$ & $[1-\nu J_K\ln\frac{W_b}{J}]^{-1}$ & $\epsilon/\epsilon_0$ at $T{=}0.5J$ \\ \hline
0.02 & 100 & 0.0105 & 11.9895 & 11.9684 & 0.999 & 1.210 & 1.117 & 1.101 & 1.623 \\
0.02 & 250 & 0.0132 & 11.9868 & 11.9604 & 0.999 & 1.215 & 1.138 & 1.124 & 1.570 \\
0.02 & 1000 & 0.0175 & 11.9825 & 11.9475 & 0.999 & 1.235 & 1.174 & 1.160 & 1.519 \\
0.05 & 100 & 0.0785 & 11.9214 & 11.7649 & 0.996 & 1.444 & 1.333 & 1.299 & 1.925 \\
0.05 & 250 & 0.1023 & 11.8976 & 11.6940 & 0.995 & 1.506 & 1.410 & 1.381 & 1.933 \\
0.05 & 1000 & 0.1443 & 11.8556 & 11.5690 & 0.993 & 1.627 & 1.547 & 1.528 & 2.018 \\
0.1 & 100 & 0.4407 & 11.5584 & 10.6946 & 0.980 & 2.026 & 1.870 & 1.854 & 2.754 \\
0.1 & 250 & 0.6289 & 11.3697 & 10.1480 & 0.971 & 2.314 & 2.168 & 2.233 & 3.053 \\
0.1 & 1000 & 1.0624 & 10.9345 & 8.9160 & 0.950 & 2.996 & 2.848 & 3.234 & 3.807 \\
0.15 & 100 & 1.5136 & 10.4764 & 7.6773 & 0.925 & 3.093 & 2.854 & 3.234 & 3.894 \\
0.15 & 250 & 2.6376 & 9.3067 & 4.8288 & 0.849 & 4.314 & 4.040 & -- & 4.619 \\
0.15 & 1000 & 5.5252 & 6.4560 & 0.1374 & 0.572 & 6.925 & 6.583 & -- & 5.733 \\
0.2 & 100 & 5.0156 & 6.9155 & 0.5503 & 0.635 & 5.765 & 5.320 & -- & 4.506 \\
0.2 & 250 & 5.7105 & 6.2838 & 0.0527 & 0.546 & 5.253 & 4.920 & -- & 4.664 \\
0.2 & 1000 & 5.9352 & 6.0645 & 0.0028 & 0.511 & 4.185 & 3.978 & -- & 4.125 \\
\end{tabular}
\end{ruledtabular}
\end{table}

\begin{table}[h!]
\footnotesize
\caption{Kondo-coupled dimer on one shared flat band of half-width $W_b$, with overlap $S$ of the two local orbitals (even and odd channels, full-density-matrix NRG with the channel-mixing exchange, $\Lambda=3$, 6000 kept states, four values of $z$, and 12000 states at $S=0$). Only entries whose spin sum rule holds to 0.01 for every $z$ are listed (28 of 57). $\xi=\epsilon/[(1-S^2)\epsilon_0]$ is the leakage factor relative to Eq.~(6) of the main text with the shared-band golden-rule width, $\xi_2$ its second-order value, and $\xi/\xi_2$ the higher-order factor. With 12000 kept states $\epsilon$ changes by 1.2\% ($S=0.6$) and 1.7\% ($S=0.9$) at $\nu J_K=0.05$, $W_b=250J$, $T=0.02J$. At $T=0.3J$ the small negative $W-W_{\rm cl}$ reflect the NRG accuracy of about 1\% of $W$.}\label{tab:l2nrg_shared}
\begin{ruledtabular}
\begin{tabular}{ccccccccccc}
$S$ & $\nu J_K$ & $W_b/J$ & $T/J$ & $\epsilon$ & $\xi$ & $\xi_2$ & $\xi/\xi_2$ & $W$ & $W_{\rm cl}$ & $W-W_{\rm cl}$ \\ \hline
0 & 0.05 & 250 & 0.02 & 0.10230 & 1.506 & 1.068 & 1.410 & 11.8976 & 11.6940 & 0.2036 \\
0 & 0.2 & 250 & 0.02 & 5.71047 & 5.253 & 1.068 & 4.920 & 6.2838 & 0.0527 & 6.2311 \\
0.3 & 0.01 & 250 & 0.02 & 0.00291 & 1.178 & 1.068 & 1.104 & 11.9971 & 11.9913 & 0.0058 \\
0.3 & 0.01 & 250 & 0.1 & 0.00294 & 1.187 & 1.076 & 1.103 & 11.9946 & 11.9892 & 0.0054 \\
0.3 & 0.01 & 250 & 0.3 & 0.00359 & 1.450 & 1.225 & 1.185 & 9.9023 & 9.9630 & -0.0607 \\
0.3 & 0.02 & 250 & 0.02 & 0.01263 & 1.277 & 1.068 & 1.196 & 11.9874 & 11.9621 & 0.0252 \\
0.3 & 0.02 & 250 & 0.3 & 0.01489 & 1.505 & 1.225 & 1.229 & 9.8196 & 9.8638 & -0.0442 \\
0.3 & 0.05 & 250 & 0.02 & 0.09643 & 1.560 & 1.068 & 1.461 & 11.9034 & 11.7115 & 0.1920 \\
0.3 & 0.05 & 250 & 0.1 & 0.09773 & 1.581 & 1.076 & 1.469 & 11.8919 & 11.7020 & 0.1899 \\
0.3 & 0.05 & 250 & 0.3 & 0.11796 & 1.908 & 1.225 & 1.558 & 9.1135 & 9.0152 & 0.0984 \\
0.3 & 0.1 & 250 & 0.02 & 0.69004 & 2.790 & 1.068 & 2.614 & 11.3060 & 9.9719 & 1.3340 \\
0.3 & 0.1 & 250 & 0.3 & 0.93696 & 3.789 & 1.225 & 3.094 & 5.1897 & 4.1229 & 1.0668 \\
0.6 & 0.01 & 250 & 0.02 & 0.00206 & 1.186 & 1.068 & 1.111 & 11.9979 & 11.9938 & 0.0041 \\
0.6 & 0.01 & 250 & 0.1 & 0.00208 & 1.195 & 1.076 & 1.110 & 11.9952 & 11.9915 & 0.0037 \\
0.6 & 0.01 & 250 & 0.3 & 0.00280 & 1.610 & 1.319 & 1.220 & 9.8436 & 9.9073 & -0.0636 \\
0.6 & 0.02 & 250 & 0.02 & 0.00897 & 1.289 & 1.068 & 1.207 & 11.9910 & 11.9731 & 0.0179 \\
0.6 & 0.02 & 250 & 0.3 & 0.01188 & 1.708 & 1.319 & 1.295 & 9.5605 & 9.6152 & -0.0547 \\
0.6 & 0.05 & 100 & 0.02 & 0.05290 & 1.520 & 1.084 & 1.403 & 11.9470 & 11.8415 & 0.1055 \\
0.6 & 0.05 & 100 & 0.3 & 0.07513 & 2.159 & 1.379 & 1.566 & 8.7749 & 8.7476 & 0.0274 \\
0.6 & 0.05 & 250 & 0.02 & 0.07408 & 1.704 & 1.068 & 1.596 & 11.9257 & 11.7782 & 0.1475 \\
0.6 & 0.05 & 250 & 0.1 & 0.07692 & 1.769 & 1.076 & 1.644 & 11.8272 & 11.6888 & 0.1383 \\
0.6 & 0.2 & 250 & 0.02 & 6.42693 & 9.238 & 1.068 & 8.653 & 4.6334 & 1.6094 & 3.0241 \\
0.9 & 0.01 & 250 & 0.02 & 0.00062 & 1.208 & 1.068 & 1.132 & 11.9994 & 11.9981 & 0.0012 \\
0.9 & 0.01 & 250 & 0.1 & 0.00063 & 1.218 & 1.077 & 1.131 & 11.9963 & 11.9955 & 0.0008 \\
0.9 & 0.01 & 250 & 0.3 & 0.00142 & 2.753 & 2.073 & 1.328 & 9.7390 & 9.8073 & -0.0683 \\
0.9 & 0.02 & 250 & 0.02 & 0.00270 & 1.307 & 1.068 & 1.224 & 11.9973 & 11.9919 & 0.0054 \\
0.9 & 0.02 & 250 & 0.3 & 0.00661 & 3.202 & 2.073 & 1.545 & 9.0749 & 9.1482 & -0.0733 \\
0.9 & 0.05 & 250 & 0.02 & 0.02891 & 2.239 & 1.068 & 2.098 & 11.9694 & 11.9125 & 0.0569 \\
\end{tabular}
\end{ruledtabular}
\end{table}

\begin{table}[h!]
\footnotesize
\caption{Two spins on one channel (overlap $S=1$) at $T=0.02J$ versus the ferromagnetic scale $x=2(W_b/J)(\nu J_K)^2$ (one-channel model of NRG Ljubljana, $\Lambda=3$, 6000 kept states, four values of $z$). $\langle\bm S_{\rm tot}^2\rangle$ is the static value $\frac32+2\langle\bm S_1\cdot\bm S_2\rangle$, and the column elastic gives its excess over the binned spectral weight of $S_{\rm tot}$, which is $\omega=0$ weight that the binned spectra do not contain. The lowest level is that of the last NRG step for every $z$. Rows with $x\ge1.25$ miss the spin sum rule by 0.05 to 0.06 for some $z$, which this elastic weight explains, so they are listed although they fail the criterion of Table~\ref{tab:l2nrg_shared}.}\label{tab:s1c}
\begin{ruledtabular}
\begin{tabular}{cccccccccc}
$x$ & $\nu J_K$ & $W_b/J$ & $\epsilon$ & $W$ & $W_{\rm cl}$ & $a$ & $\langle\bm S_{\rm tot}^2\rangle$ & elastic & lowest level \\ \hline
0.002 & 0.01 & 10 & 0.00000 & 12.0000 & 12.0000 & 1.0000 & 0.0000 & 0.0000 & singlet \\
0.005 & 0.01 & 25 & 0.00000 & 12.0000 & 12.0000 & 1.0000 & 0.0000 & 0.0000 & singlet \\
0.02 & 0.01 & 100 & 0.00000 & 12.0000 & 12.0000 & 1.0000 & 0.0000 & 0.0000 & singlet \\
0.05 & 0.01 & 250 & 0.00000 & 12.0000 & 12.0000 & 1.0000 & 0.0000 & 0.0000 & singlet \\
0.05 & 0.05 & 10 & 0.00000 & 12.0000 & 12.0000 & 1.0000 & 0.0000 & 0.0000 & singlet \\
0.125 & 0.05 & 25 & 0.00000 & 12.0000 & 12.0000 & 1.0000 & 0.0000 & 0.0000 & singlet \\
0.2 & 0.1 & 10 & 0.00000 & 12.0000 & 12.0000 & 1.0000 & 0.0000 & 0.0000 & singlet \\
0.5 & 0.1 & 25 & 0.00000 & 12.0000 & 12.0000 & 1.0000 & 0.0000 & 0.0000 & singlet \\
0.5 & 0.05 & 100 & 0.00000 & 12.0000 & 12.0000 & 1.0000 & 0.0000 & 0.0000 & singlet \\
1.25 & 0.05 & 250 & 0.12250 & 5.7646 & 4.8837 & 0.6731 & 0.6537 & 0.3445 & singlet \\
2 & 0.1 & 100 & 2.00413 & 4.0000 & 4.0000 & 0.0000 & 2.0000 & 0.6554 & triplet \\
5 & 0.1 & 250 & 2.46976 & 4.0000 & 4.0000 & 0.0000 & 2.0000 & 0.5881 & triplet \\
\end{tabular}
\end{ruledtabular}
\end{table}

\subsection{Heisenberg chains and rings}
For open and periodic Heisenberg chains of $N=4$--$12$ spins $\frac12$ we computed the exact spatial QFI matrix, synthetic spectra of every site ($r=0.5$, $\eta=0.5\%$) and the certified local bounds (Table~\ref{tab:s13}). No certified local bound exceeded the exact $D_i$ in 50 cases, and $\langle D_i\low/D_i\rangle=0.61$--$0.72$, close to the value $0.75$ of Eq.~(4) of the main text at low temperature. The largest eigenvalue of $K$, however, is only weakly constrained by local information for $N>2$: even with the exact $D_i$, Eq.~\eqref{eq:cluster} and the semidefinite program give $1.7$--$4.0$ against true values $3.1$--$11.8$, because for $N>2$ total-spin conservation fixes only the row sums of $K$ and not its individual off-diagonal elements. With certified local bounds the product threshold $\lambda\low>2$ is reached below $T\approx0.1$--$0.3J$, depending on $N$ and boundary conditions, and the private-environment threshold is not reached. Stronger statements for longer clusters therefore require intersite information beyond local spectra.

\begin{table}[h!]
\scriptsize
\caption{Heisenberg chains and rings of $N$ spins $\frac12$ (exact diagonalization). $\lambda_{\max}(K)$ is the largest eigenvalue of the exact spatial QFI matrix. The bounds from local information use Eq.~\eqref{eq:cluster} or semidefinite programming (SDP) with $\epsilon=0$, either with the exact local QFIs or with the certified local bounds from synthetic spectra ($r=0.5$, $\eta=0.5\%$, $n_b=181$, joint confidence level 95\%). No certified local bound exceeded the exact $D_i$.}\label{tab:s13}
\begin{ruledtabular}
\begin{tabular}{ccccccccc}
$N$ & boundary & $T/J$ & $\lambda_{\max}(K)$ & Eq.~\\eqref{eq:cluster}, exact $D_i$ & SDP, exact $D_i$ & Eq.~\\eqref{eq:cluster}, certified & SDP, certified & $\langle D_i\low/D_i\rangle$ \\ \hline
4 & open & 0.05 & 7.464 & 4.000 & 4.000 & 2.898 & 2.898 & 0.725 \\
4 & open & 0.1 & 7.421 & 3.984 & 3.991 & 2.876 & 2.882 & 0.722 \\
4 & open & 0.2 & 6.429 & 3.608 & 3.757 & 2.496 & 2.620 & 0.692 \\
4 & open & 0.3 & 4.954 & 2.975 & 3.305 & 1.974 & 2.228 & 0.662 \\
4 & open & 0.5 & 3.068 & 1.937 & 2.343 & 1.193 & 1.502 & 0.609 \\
6 & open & 0.05 & 8.482 & 3.600 & 3.600 & 2.604 & 2.605 & 0.724 \\
6 & open & 0.1 & 8.259 & 3.550 & 3.575 & 2.541 & 2.564 & 0.716 \\
6 & open & 0.2 & 6.440 & 3.112 & 3.332 & 2.130 & 2.322 & 0.684 \\
6 & open & 0.3 & 4.906 & 2.631 & 2.972 & 1.752 & 2.025 & 0.664 \\
6 & open & 0.5 & 3.167 & 1.809 & 2.127 & 1.123 & 1.371 & 0.616 \\
8 & open & 0.05 & 9.264 & 3.427 & 3.428 & 2.476 & 2.477 & 0.723 \\
8 & open & 0.1 & 8.692 & 3.335 & 3.383 & 2.366 & 2.416 & 0.709 \\
8 & open & 0.2 & 6.352 & 2.907 & 3.148 & 1.991 & 2.201 & 0.684 \\
8 & open & 0.3 & 4.914 & 2.508 & 2.826 & 1.675 & 1.930 & 0.667 \\
8 & open & 0.5 & 3.223 & 1.760 & 2.026 & 1.095 & 1.303 & 0.618 \\
10 & open & 0.05 & 9.881 & 3.328 & 3.331 & 2.399 & 2.404 & 0.721 \\
10 & open & 0.1 & 8.858 & 3.199 & 3.269 & 2.256 & 2.331 & 0.705 \\
10 & open & 0.2 & 6.300 & 2.805 & 3.051 & 1.925 & 2.136 & 0.686 \\
10 & open & 0.3 & 4.945 & 2.448 & 2.746 & 1.636 & 1.876 & 0.667 \\
10 & open & 0.5 & 3.257 & 1.733 & 1.970 & 1.079 & 1.266 & 0.619 \\
12 & open & 0.05 & 10.365 & 3.261 & 3.267 & 2.345 & 2.355 & 0.719 \\
12 & open & 0.1 & 8.879 & 3.105 & 3.195 & 2.184 & 2.278 & 0.703 \\
12 & open & 0.2 & 6.286 & 2.747 & 2.992 & 1.887 & 2.095 & 0.687 \\
12 & open & 0.3 & 4.976 & 2.411 & 2.696 & 1.612 & 1.841 & 0.667 \\
12 & open & 0.5 & 3.278 & 1.716 & 1.934 & 1.068 & 1.241 & 0.620 \\
4 & periodic & 0.05 & 8.000 & 4.000 & 4.000 & 2.898 & 2.898 & 0.725 \\
4 & periodic & 0.1 & 7.998 & 3.999 & 3.999 & 2.897 & 2.897 & 0.724 \\
4 & periodic & 0.2 & 7.748 & 3.915 & 3.915 & 2.796 & 2.796 & 0.714 \\
4 & periodic & 0.3 & 6.754 & 3.563 & 3.563 & 2.453 & 2.453 & 0.688 \\
4 & periodic & 0.5 & 4.234 & 2.510 & 2.510 & 1.617 & 1.617 & 0.644 \\
6 & periodic & 0.05 & 9.328 & 3.600 & 3.600 & 2.606 & 2.606 & 0.724 \\
6 & periodic & 0.1 & 9.285 & 3.592 & 3.592 & 2.593 & 2.593 & 0.722 \\
6 & periodic & 0.2 & 8.133 & 3.361 & 3.361 & 2.339 & 2.339 & 0.696 \\
6 & periodic & 0.3 & 6.190 & 2.914 & 2.914 & 1.952 & 1.952 & 0.670 \\
6 & periodic & 0.5 & 3.582 & 2.027 & 2.027 & 1.281 & 1.281 & 0.632 \\
8 & periodic & 0.05 & 10.324 & 3.428 & 3.428 & 2.479 & 2.479 & 0.723 \\
8 & periodic & 0.1 & 10.103 & 3.398 & 3.398 & 2.436 & 2.436 & 0.717 \\
8 & periodic & 0.2 & 7.831 & 3.070 & 3.070 & 2.106 & 2.106 & 0.686 \\
8 & periodic & 0.3 & 5.648 & 2.652 & 2.652 & 1.772 & 1.772 & 0.668 \\
8 & periodic & 0.5 & 3.397 & 1.885 & 1.885 & 1.186 & 1.186 & 0.629 \\
10 & periodic & 0.05 & 11.116 & 3.332 & 3.332 & 2.407 & 2.407 & 0.722 \\
10 & periodic & 0.1 & 10.521 & 3.270 & 3.270 & 2.324 & 2.324 & 0.711 \\
10 & periodic & 0.2 & 7.395 & 2.904 & 2.904 & 1.987 & 1.987 & 0.684 \\
10 & periodic & 0.3 & 5.349 & 2.531 & 2.531 & 1.693 & 1.693 & 0.669 \\
10 & periodic & 0.5 & 3.352 & 1.824 & 1.824 & 1.146 & 1.146 & 0.628 \\
12 & periodic & 0.05 & 11.759 & 3.270 & 3.270 & 2.358 & 2.358 & 0.721 \\
12 & periodic & 0.1 & 10.627 & 3.173 & 3.173 & 2.239 & 2.239 & 0.706 \\
12 & periodic & 0.2 & 7.042 & 2.807 & 2.807 & 1.923 & 1.923 & 0.685 \\
12 & periodic & 0.3 & 5.211 & 2.467 & 2.467 & 1.651 & 1.651 & 0.669 \\
12 & periodic & 0.5 & 3.341 & 1.789 & 1.789 & 1.122 & 1.122 & 0.627 \\
\end{tabular}
\end{ruledtabular}
\end{table}

\end{document}